\documentclass[twocolumn,twocolappendix]{aastex631}

\usepackage{amsmath}
\usepackage{lipsum}
\usepackage{float}
\usepackage{afterpage}
\usepackage{makecell}
\usepackage{upgreek}
\usepackage{bm}
\usepackage{ulem}
\usepackage{starfont}
\usepackage[T1]{fontenc}
\usepackage{calligra}

\newcommand\as{\bgroup\markoverwith{\textcolor[rgb]{.5, 0, .6}{\rule[0.5ex]{8pt}{1.5pt}}}\ULon}

\newcommand{\be}{\begin{eqnarray}}
\newcommand{\ee}{\end{eqnarray}}
\renewcommand{\vec}[1]{\mathbf{#1}}

\newcommand{\RN}[1]{\textup{\uppercase\expandafter{\romannumeral#1}}}

\DeclareMathAlphabet{\mathcalligra}{T1}{calligra}{m}{n}
\DeclareFontShape{T1}{calligra}{m}{n}{<->s*[2.2]callig15}{}

\graphicspath{{./}{figures/}}
\shorttitle{}
\shortauthors{Dittmann, Dempsey, \& Li}

\begin{document}

\title{The Evolution of Inclined Binary Black Holes in the Disks of Active Galactic Nuclei}
\shorttitle{Inclined Binaries in AGN Disks}

\correspondingauthor{Alexander J. Dittmann}
\email{dittmann@astro.umd.edu}

\author[0000-0001-6157-6722]{Alexander J. Dittmann}
\affil{Theoretical Division, Los Alamos National Laboratory, Los Alamos, NM 87545, USA}
\affil{Department of Astronomy and Joint Space-Science Institute, University of Maryland, College Park, MD 20742-2421, USA}

\author[0000-0001-8291-2625]{Adam M. Dempsey}
\affil{X-Computational Physics Division, Los Alamos National Laboratory, Los Alamos, NM 87545, USA}

\author[0000-0003-3556-6568]{Hui Li}
\affil{Theoretical Division, Los Alamos National Laboratory, Los Alamos, NM 87545, USA}

\begin{abstract} 
The accretion disks that fuel active galactic nuclei (AGN) may house numerous stars and compact objects, formed in situ or captured from nearby star clusters. Embedded neutron stars and black holes may form binaries and eventually merge, emitting gravitational waves detectable by LIGO/VIRGO. AGN disks are a particularly promising environment for the production of high-mass gravitational wave events involving black holes in the pair instability mass gap, and may facilitate electromagnetic counterparts to black hole binary mergers. However, many orders of magnitude separate the typical length scales of binary formation and those on which gravitational waves can drive binary inspirals, making binary mergers inside the disk uncertain. Previous hydrodynamical simulations of binaries have either been restricted to two dimensions entirely, or focused on binaries aligned with the midplane of the disk. Herein we present the first three-dimensional, high-resolution, local shearing-box hydrodynamical simulations of disk-embedded binaries over a range of orbital inclinations. 
We find that retrograde binaries can shrink up to four times as quickly as prograde binaries, and that 
all binaries not perfectly aligned (or anti-aligned) with the AGN disk are driven into alignment.
An important consequence of this is that initially retrograde binaries will traverse the inclinations where von Zeipel-Lidov-Kozai oscillations can drive binary eccentricities to large values, potentially facilitating mergers. 
We also find that interactions with the AGN disk may excite eccentricities in retrograde binaries and cause the orbits of embedded binaries to precess. 
\end{abstract}

\keywords{Astrophysical fluid dynamics (101); Active galactic nuclei (16); Black holes (162); Accretion (14); Gravitational wave sources (677)}

\section{Introduction}
The accretion disks around supermassive black holes (SMBH), which power active galactic nuclei (AGN) \citep[e.g.][]{1969Natur.223..690L,2008ARA&A..46..475H}, are a potential host site for merging stellar-mass black holes. 
Stellar mass black holes embedded in these disks will grow via gas accretion, and potentially through repeated mergers facilitated by the deep potential well of the SMBH \citep[e.g.][]{2012MNRAS.425..460M,2021NatAs...5..749G}. 
Thus, AGN disks are readily able to produce black holes within the pair instability mass gap, such as those inferred to have caused GW190521 ($85_{-14}^{+21}\,M_\odot$ and $66_{-18}^{+17}\,M_\odot$ at 90\% confidence, \citet{2020PhRvL.125j1102A}; see however \citet{2021ApJ...907L...9N}).
Additionally, the presence of gas introduces the \textit{possibility} of electromagnetic counterparts to stellar-mass black hole binary mergers \citep[e.g.][]{2017MNRAS.464..946S,2019ApJ...884L..50M,2023ApJ...950...13T}. 
Flaring AGN have been associated with the localization error volumes of not only GW190521 \citep{2020PhRvL.124y1102G}, but also eight other gravitational wave events \citep{2023ApJ...942...99G} during the third observing run of Advanced LIGO and Advanced VIRGO \citep{2021arXiv211103606T}, although these associations may not be statistically robust \citep[e.g.][]{2021ApJ...907L...9N,2021ApJ...914L..34P}. 

Numerous investigations have estimated the rates of black hole binary mergers within AGN disks using semi-analytical methods or one-dimensional simulations \citep[e.g.][]{2017MNRAS.464..946S,2017ApJ...835..165B,2020ApJ...898...25T,2020MNRAS.494.1203M,2020MNRAS.498.4088M}, and it has been estimated that up to $\sim80$\% of the black hole binary mergers detected by LIGO/VIRGO may occur within AGN disks \citep{2022MNRAS.tmp.2697S}.

The progenitors of these GW events -- stars and compact objects -- may be captured into the disk from a nuclear star cluster through, for example, torques from the gaseous disk or dynamical friction \citep[e.g.][]{1991MNRAS.250..505S,1993ApJ...409..592A,1995MNRAS.275..628R,2020ApJ...889...94M}. Additionally, gravitational instability in the outer regions of AGN disks may lead to star formation \citep[e.g.][]{1980SvAL....6..357K,2003MNRAS.339..937G}. Furthermore, stellar evolution simulations suggest that most stars captured into or formed within AGN disks will leave behind black holes and neutron stars at the end of their lives \citep{2021ApJ...910...94C,2021ApJ...914..105J,2021ApJ...916...48D,2023ApJ...946...56D,2023MNRAS.tmp.2707A}, and neutron stars within the disk will likely accrete from the disk to the point of collapse \citep[e.g.][]{2021ApJ...915...10P,2021ApJ...923..173P}. 
Therefore, it is plausible that AGN disks host a substantial population of stellar-mass black holes.\footnote{Although this work is generally focused on binary black holes embedded in AGN disks, some aspects of our results are also applicable to embedded stellar binaries, \citep[e.g.][]{2022ApJ...929..133J}.}

A growing body of work has studied binary formation within AGN disks using multidimensional N-body simulations \citep[e.g.][]{2019ApJ...878...85S,2020ApJ...903..133S,2022ApJ...934..154L} as well as hydrodynamical simulations \citep[e.g.][]{2023ApJ...944L..42L,2023MNRAS.524.2770R,2023arXiv230911561W,2023arXiv230914433R}. 
An assumption that pervades the literature on black hole binary mergers within AGN disks is that gas-binary interactions help drive binaries towards inspiral.
This assumption stems from early analytical \citep[e.g.][]{1991MNRAS.248..754P,1991ApJ...370L..35A} and numerical \citep[e.g.][]{2008ApJ...672...83M} work on isolated binary-disk interaction. 
However, subsequent higher-resolution and longer-duration simulations of circumbinary accretion have demonstrated that binary inspirals are far from universal \citep[e.g.][see also the recent review \citealt{2023ARA&A..61..517L}]{2019ApJ...875...66M,2019ApJ...871...84M,2020ApJ...889..114M,2020ApJ...900...43T,2021ApJ...921...71D,2022MNRAS.513.6158D,2023MNRAS.tmp.2762W}. 

Whether inspiral or outspiral due to interactions with a circumbinary disk depends on the binary eccentricity \citep[e.g.][]{2017MNRAS.466.1170M}, disk aspect ratio \citep[e.g.][]{2020ApJ...900...43T,2022MNRAS.513.6158D}, disk viscosity \citep[e.g.][]{2017MNRAS.466.1170M,2022MNRAS.513.6158D}, and binary mass ratio \citep[e.g.][]{2020ApJ...889..114M,2021ApJ...918L..36D}.

Studies of viscous circumbinary disks inclined relative to their binaries have found that the binary and disk are typically brought into alignment in the circular case \citep[e.g.][]{2015ApJ...800...96L,2019ApJ...875...66M}, but also that disks may equilibriate to a polar configuration about eccentric binaries \citep[e.g.][]{2017ApJ...835L..28M}. 

Compared to the study of isolated binaries, the study of embedded binaries is nascent. 
Early studies pursued global two-dimensional simulations,
identifying the importance of resolution \citep[c.f.][]{2011ApJ...726...28B,2021ApJ...911..124L}, binary orientation \citep{2021ApJ...911..124L}, and disk thermodynamics \citep{2022ApJ...928L..19L}.
More recently, two-dimensional shearing-sheet simulations have surveyed binary eccentricity and mass ratio \citep{2022MNRAS.517.1602L}, typically finding that binaries shrink and eccentricities are damped. Recent two-dimensional studies have also surveyed the gas equation of state and binary separation \citep{2023MNRAS.522.1881L}. 

Heretofore the only three-dimensional simulations of binary stellar-mass black holes embedded in AGN disks have been those presented in \citet{2022ApJ...940..155D}. Using isothermal shearing-box simulations that directly evolve the three-body orbital evolution of the system, that work illustrated the dependence of binary orbital evolution on the binary separation; balance between gravitational forces and gas pressure; and dimensionality of the problem, finding that three dimensional flows dramatically differ in structure compared to those found in two-dimensional simulations.

All previous simulations of binaries embedded in AGN disks have focused on binary orbits aligned or anti-aligned with the midplane of the SMBH disk, while in reality binaries may form with a wide range of inclinations and eccentricities. In particular, the two-dimensional simulations of gas-assisted binary formation by \citet{2023ApJ...944L..42L} suggest that newly-formed binaries should form with high eccentricities, predominantly in retrograde configurations.
Thus motivated, we focus in the present study on the orbital evolution of inclined
embedded binaries, and will present the evolution of eccentric binaries in a companion paper, in both cases investigating both prograde and retrograde binaries.

We review AGN disk models and the orbits of embedded binaries in Section \ref{sec:basic}, and describe our simulation methodology in Section \ref{sec:methods}. We present the results of our simulations in Section \ref{sec:results}, including that retrograde binaries can inspiral more than four times faster than prograde binaries and binary inclinations are gradually driven to zero. We discuss some shortcomings of the present study, as well as its implications for mergers in AGN disks and various dynamical phenomena, in Section \ref{sec:discussion}. We conclude in Section \ref{sec:conclusion}, and provide additional technical details in Appendix \ref{sec:appendix}. We present some convergence tests in Appendix \ref{sec:appendixSoft} and tabulate some of our results in Appendix \ref{sec:appendixTable}.

\section{Binaries in AGN Disks}\label{sec:basic}
\subsection{Objects embedded in AGN disks}\label{sec:disks}
The accretion disks that fuel AGN are typically modeled by finding solutions to the equations of viscous hydrodynamics, often under the assumptions of azimuthal symmetry, a steady state, and the limit of a thin disk such that the disk pressure scale height $H$ at a given radial distance $r_\bullet$ is $H\ll r_\bullet$ \citep[e.g.][]{1973A&A....24..337S,1973blho.conf..343N,2002apa..book.....F,2003MNRAS.341..501S,2005ApJ...630..167T}, although models which relax some of these assumptions have also been developed \citep[e.g.][]{1988ApJ...332..646A,2009ApJS..183..171S,2022ApJ...928..191G}. A common feature of many disk models is that at large distances from the center the free-fall timescale of the gas disk becomes shorter than the dynamical timescale leading to gravitational instability \citep[e.g.][]{1964ApJ...139.1217T} and the formation of stars \citep[e.g.][]{1980SvAL....6..357K,2023MNRAS.521.4522D}. Some disk models have accounted for this by imposing marginal stability arbitrarily \citep[e.g.][]{2003MNRAS.341..501S} or by invoking feedback from stars or black holes embedded in the disk \citep[e.g.][]{2005ApJ...630..167T,2020MNRAS.493.3732D,2022ApJ...928..191G}.

\begin{figure}
\includegraphics[width=\columnwidth]{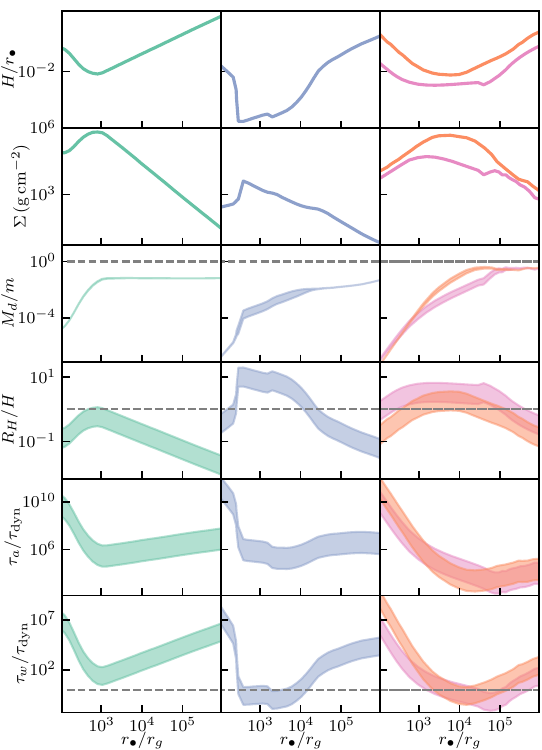}
\caption{Four example AGN disk models: the left and center panels display AGN disk models (those of \citet{2003MNRAS.341..501S} and \citet{2005ApJ...630..167T} respectively) around $M_\bullet=10^8\,M_\odot$ SMBHs computed in \citet{2016ApJ...819L..17B}; and the right column displays models for disks around $4\times10^6\,M_\odot$ SMBHs computed in \citet{2020MNRAS.493.3732D}, based on \citet{2005ApJ...630..167T} but making different assumptions about feedback and gas opacity.
The first two rows display the disk aspect ratio and surface density.
The bottom four rows display the range of $M_d/m$, $R_H/H$, $\tau_a/\tau_{\rm orb}$, and $\tau_w/\tau_{\rm orb}$ for objects with masses ranging from $5\,M_\odot$ to $300\,M_\odot$. Horizontal dashed gray lines plot the line $y=1$ where applicable.}
\label{fig:disks}
\end{figure}

We provide examples of several AGN disk models from the literature in Figure \ref{fig:disks} (see also Figure 1 of \citealt{2022ApJ...940..155D}), specifically their aspect ratios $H/r_\bullet$, surface densities ($\Sigma$), the ratio of local disk mass to object mass, ratio of the Hill radius to disk scale height,
and characteristic timescales for stellar mass objects to change their orbits.

Restricting ourselves to thin disks $(H\ll r_\bullet)$ which are vertically isothermal, the density profile as a function of height above the disk midplane ($z$) at a given radius is $\rho(z)=\rho_0\exp{[-(z/H)^2/2]}$, where $\rho_0$ is the density in the disk midplane and 
$H$ is the pressure scale height of the disk. Furthermore, $H=c_s/\Omega$, where $c_s$ is the sound speed within the disk at a given radius, $\Omega=\sqrt{GM_\bullet /r_\bullet^{3}}$ is the angular velocity of the disk, and $M_\bullet$ is the SMBH mass.
Considering objects\footnote{Within this subsection the difference between individual objects and tightly-bound binaries is immaterial so we use `object' in place of either.} of mass $m\ll M_\bullet$ embedded within the AGN disk, the characteristic length scale within which gas is bound to the embedded object rather than the SMBH is the Hill radius \citep{hill1878researches}, 
\begin{equation}
R_H = r_\bullet\left(\frac{m}{3M_\bullet}\right)^{1/3}.
\end{equation}
The local disk mass ($M_d$) within the Hill sphere of an object is given, assuming only vertical density variation, by
\begin{equation}
M_d\!=\!2\pi\rho_0 H\!\!\left( \!\sqrt{\frac{\pi}{2}}(R_H^2\!-\!H^2){\rm erf}\!\left(\!\frac{R_H}{H\sqrt{2}}\right)\!\!+R_HHe^{-\frac{R_H^2}{2H^2}}\right)\!\!.
\end{equation}

The ratio $R_H/H$ is a measure of the strength of gravitational forces relative to gas pressure, as well as the degree to which the disk is vertically stratified within the vicinity of the embedded object. In the limit $R_H\ll H$, the gas surrounding an embedded object has a near-constant density and the local disk mass is $M_d\approx4\pi\rho_0R_H^3/3$. In the limit $R_H\gg H$, the gas about an embedded object is strongly stratified and the local disk mass is $M_d\approx\sqrt{2\pi^3}\rho_0R_H^2H$. As illustrated in Figure \ref{fig:disks}, objects embedded in AGN disks can occupy both extremes. Previous studies have demonstrated that $R_H/H$ significantly influences the orbital evolution of binaries embedded within accretion disks, with more-embedded binaries inspiraling more quickly at a given orbital separation \citep[e.g.][]{2022ApJ...940..155D}.

The ratio $M_d/m$ gauges whether the self-gravity of the gas disk is dynamically relevant to the orbital evolution of embedded objects. Throughout most of the AGN disk models shown in Figure \ref{fig:disks} the self-gravity of the AGN disk should have a negligible effect on the dynamics of embedded objects, although it may be marginally important in the outer regions of the disks modeled in \citet{2020MNRAS.493.3732D}. Therefore, we are somewhat justified neglecting in our simulations the disk self-gravity and the back-reaction of gas upon the black holes.

The orbit of an embedded object about the SMBH can be characterized by a semi-major axis $a$, an eccentricity $e$, and an inclination angle relative to the disk midplane $i$. As embedded objects orbit within the disk, they excite waves that can transfer orbital energy and angular momentum between the disk and object. 
Because the mass of an embedded object is typically much smaller than the mass of the corresponding SMBH, we approximate the timescales to change the orbital elements $\tau_a$ for the semi-major axis and $\tau_w$ for the eccentricity and inclination) using the classical Type I values
\citep{1980ApJ...241..425G,2002ApJ...565.1257T,2004ApJ...602..388T}
\be
\tau_a &\approx& \left(\frac{M_\bullet}{\Sigma r_\bullet^2}\right)\left(\frac{H}{r_\bullet}\right)^2q^{-1}\tau_{\rm orb}, \label{eq:taua} \\
\tau_w &\approx& \left(\frac{H}{r_\bullet}\right)^2 \tau_a \label{eq:tauw}
\ee
where $q\equiv m/M_\bullet$ and $\tau_{\rm orb}\equiv\Omega^{-1}$ is the local orbital timescale.
Because astrophysical disks generally have $H/r_\bullet<1$, $\tau_w$ will generally be shorter than $\tau_a$.

The ratio between $\tau_a$ and $\tau_{\rm orb}$ measures the rate of change of the semi-major axis of an embedded object. If $\tau_a/\tau_{\rm orb}\gg1,$ then the physical properties of the AGN disk are roughly constant on the orbital timescale of the embedded object. In this case the local AGN disk properties in the vicinity of an embedded objects are approximately constant, whereas if $\tau_a\sim\tau_{\rm orb}$ the disk properties change relatively rapidly. In our simulations of embedded binaries, we assume that the properties of the AGN disk are constant in time, which is justified when $\tau_a/\tau_{\rm orb}\gg1$. Considering the disk models shown in Figure \ref{fig:disks}, the steady-state assumption in our simulations is reasonable in all AGN disk regions save for the outer regions of the disk models calculated in \citet{2020MNRAS.493.3732D}. 

Objects formed within the disk will typically have inclinations $i\lesssim H/r_\bullet$, while objects captured into the disk may have larger initial inclinations ($i>H/r_\bullet$) that are eventually damped to $i\lesssim H/r_\bullet$.
Additionally, gravitational scattering events between embedded objects may excite orbital eccentricities and inclinations over time \citep[e.g.][]{2000Icar..143...28S}. For example, \citet{2017MNRAS.464..946S} estimated that over $10^8$ years, typical inclinations exceeded $H/r$ by factors of order unity far from the SMBH and more than an order of magnitude closer in.
Assuming that binaries form within the disk with orbital separations $a_b\sim R_H$ \citep[e.g.,][]{2023ApJ...944L..42L}, in regions of the disk where $R_H/H\lesssim1$ and $\tau_w/\tau_{\rm orb}\gtrsim1$ binaries may form with virtually any orientation relative to the disk midplane. On the other hand, where $\tau_w/\tau_{\rm orb}\lesssim1$, binaries will likely form in the disk midplane, either aligned or anti-aligned with the orbit of the binary about the SMBH, although the linear analysis performed in \citet{2004ApJ...602..388T} breaks down in the $\tau_w/\tau_{\rm orb}\lesssim1$ limit. In regions with $\tau_w/\tau_{\rm orb}>1$, binary-single interactions may excite inclination even in binaries which are initially coplanar with the disk midplane \citep[e.g.][]{2020ApJ...898...25T}. 

\subsection{Binary Orbits in AGN disks}
In this section, we introduce basic definitions related to black hole binaries (BHBs) in AGN disks, drawing from Chapter 6 of \citet{1988fcm..book.....D}. Subsequently, we briefly review relevant dynamical phenomena. 

\subsubsection{Definitions}\label{sec:def}
The BHB is characterized by both its specific energy and its specific angular momentum vector, 
\be
\mathcal{E} = \frac{1}{2} \vec{v} \cdot \vec{v} - \frac{G m_b}{|\vec{r}|} , \qquad \vec{h} = \vec{r} \times \vec{v} ,
\ee
where $m_b = m_1 + m_2$ is the binary mass, $\vec{r} = \vec{r}_2 - \vec{r}_1$ is the relative position vector and $\vec{v} = \vec{v}_2 - \vec{v}_1$ is the relative velocity vector.  
From $\mathcal{E}$ and $h \equiv \sqrt{ \vec{h} \cdot \vec{h}}$, we define the orbital semi-major axis, $a_b$, and eccentricity, $e_b$, using the well-known relations
\be
a_b = -\frac{G m_b}{2 \mathcal{E}} , \qquad e_b^2 = 1 - \frac{h^2}{G m_b a_b} . 
\ee
In addition to the scalar eccentricity, we also define the eccentricity vector which, in the orbital plane, points toward the periapse,
\be \label{eq:ecc}
\mathbf{e}=\frac{\mathbf{v}\times\mathbf{h}}{Gm_b}-\frac{\mathbf{r}}{|\mathbf{r}|} . 
\ee

\begin{figure}
\centering
\includegraphics[trim={1.6cm 0cm 1.6cm 0cm},clip, width=\columnwidth]{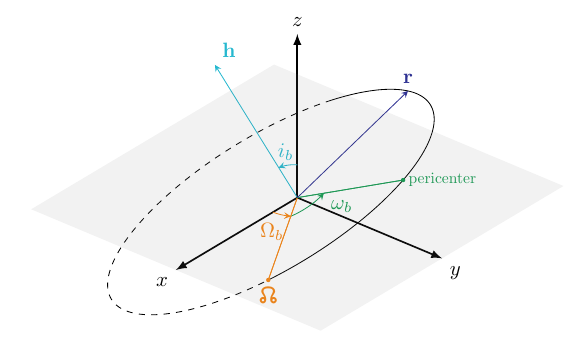}
\caption{A schematic of a binary orbit oriented in three-dimensional space. The reference plane is shown in gray, and the portion of the orbit below the plane is plotted using a dashed line. The ascending node, the point at which $\mathbf{r}$ passes from below to above the reference plane, is marked as \NorthNode
.}\label{fig:orbit}
\end{figure}

Orienting an inclined and eccentric BHB in space requires three additional angles $(i_b, \omega_b, \Omega_b)$ which are respectively the orbit inclination, argument of periapse, and longitude of the ascending node.
For reference, we illustrate this connection in Figure \ref{fig:orbit}.
From the schematic, the orientation angles are related to $\vec{h}$ and $\vec{e}$ via, 
\be
\cos{i_b} &=& h_z / h, \label{eq:incdef}  \\ 
\cos{\Omega_b} &=&-h_y/h_\perp, \label{eq:cOmegadef} \\
\sin{\Omega_b} &=&h_x/h_\perp, \label{eq:sOmegadef} \\
e_b\cos{\omega_b} &=& e_x\cos{\Omega_b} + e_y\sin{\Omega_b}, \label{eq:comegadef}\\
e_b\sin{\omega_b} &=& -e_x\sin{\Omega_b} + e_y\cos{\Omega_b}, \label{eq:somegadef}
\ee
where because $\omega_b$ and $\Omega_b$ are each confined to a plane they can be oriented without ambiguity, $i_b$ simply takes values between $0$ and $\pi$,\footnote{A number of textbooks on dynamics have included typos in their analogs to these expressions. See, for example, the list of known errors in \citet[][\url{https://www.solarsystemdynamics.info/known-errors}]{1999ssd..book.....M}.} and we have defined the magnitude of the in-midplane component of the binary angular momentum vector to be $h_\perp\equiv\sqrt{h_x^2+h_y^2}$. We note that the role of $\Omega_b$ in the expressions for $\omega_b$ is simply a rotation in the $x-y$ plane, and that $\Omega_b$ is undefined when $h_\perp=0$, in which case only $\Omega_b+\omega_b$ is a meaningful quantity and we set $\Omega_b=0$. 

Gravitational interaction with the surrounding gas changes the binary's total energy, $E$, and angular momentum, $\vec{J}$. 
These are related to the specific values through the reduced mass $\mu_b = m_1 m_2 /m_b$ such that,
\be
E \equiv \mu_b \mathcal{E} , \qquad \vec{J} = \mu_b \vec{h}
\ee
The power delivered to the binary, $\dot{E}$, and the total torque on the binary, $\dot{\vec{J}}$, together with the mass accretion rates $\dot{m}_{1,2}$, induce changes in $a_b$ and $e_b$ at the rates,
\be
\frac{\dot{a}_b}{a_b} = \frac{\dot{m}_1}{m_1} + \frac{\dot{m}_2}{m_2} - \frac{\dot{E}}{E} \label{eq:abdot}\\ 
\frac{e_b \dot{e}_b}{1 - e_b^2} =  \frac{\dot{m}_b}{m_b} + \frac{3\dot{\mu}_b}{2\mu_b}- \frac{\dot{J}}{J} - \frac{\dot{E}}{2E}, \label{eq:ebdot}
\ee 
where $\dot{J}=\dot{\vec{J}}\cdot\vec{J}/J.$ 
In addition to $\dot{a}_b$ and $\dot{e}_b$, the orientation angles change with the rates

\be
\dot{i}_b &=& \frac{\dot{h}_zh-h_z\dot{h}}{h\sqrt{h^2-h_z^2}}\label{eq:idot}, \\
\dot{\Omega}_b &=& \frac{\dot{
h}_yh_x-\dot{h}_xh_y}{h_x^2+h_y^2},\label{eq:Omdot}
\ee
\begin{flalign}
\dot{\omega}_b = \frac{\dot{e}_b\!\cos{\!\omega_b\!}\!-\!(\dot{e}_x\!+\!e_y\dot{\Omega}_b\!)\!\cos{\!\Omega_b\!}\!-\!(\dot{e}_y\!-\!e_x\dot{\Omega}_b\!)\!\sin{\!\Omega_b\!}}{e_y\cos{\!\Omega_b}\!-\!e_x\sin{\Omega_b}},\label{eq:omegadot}
\end{flalign}
which reduces, when $\Omega_b=\dot{\Omega}_b=0$, to the familiar expression 
\begin{equation}
\dot{\omega}_b= \frac{\dot{e}_ye_x- \dot{e}_xe_y}{e_b^2}.
\end{equation}

In Appendix \ref{sec:appendix} we provide further details on how we measure changes in the binary energy, angular momentum, and other quantities; how we derive average rates of change of binary orbital elements using these measurements; and list additional mathematical relations omitted here for brevity.
In Section \ref{sec:results}, we will make use of a coordinate system aligned with the BHB made up of the triad $\vec{h}, \vec{r}$, and $\vec{h}\times \vec{r}$. 

\subsubsection{Dynamics}
We briefly draw attention to a few flavors of dynamical interactions between binaries, such as those described above, and a distant, perturbing third body. First, binaries with large separations ( $a_b\gtrsim0.4\,R_H$ ) are known to evolve chaotically; tidal forces from the SMBH are likely to disrupt such loosely bound binaries \citep[e.g.][]{1995ApJ...455..640E,2001MNRAS.321..398M}. Another process that may operate, notable for functioning even in purely coplanar triple systems, is the evection resonance \citep[][]{1994AJ....108.1943T}. This process relies on the orbital angular frequency of the outer black hole binary about the SMBH ($n_\bullet$) being commensurable to the rate of change of the longitude of periapsis $n_\bullet\approx\dot{\varpi}_b$, where $\varpi_b=\omega_b + \Omega_b$. The resonance may be able to drive black hole binaries migrating inward through AGN disks towards high eccentricities \citep[and thus faster gravitational wave-driven mergers, e.g.][]{1964PhRv..136.1224P}, although this mechanism is thought to preferentially apply to binaries in AGN disks composed of an intermediate-mass black hole and a stellar-mass black hole \citep{2022arXiv220406002M,2022ApJ...934..141B}.
If binaries acquire appreciable eccentricities through the evection resonance or other processes, their inclination may also be excited through the eviction resonance \citep{1994AJ....108.1943T} at the expense of the eccentricity of the binary.

The von Zeipel-Lidov-Kozai (ZLK) effect \citep[][see also, e.g., \citealt{2019MEEP....7....1I} for a review]{1910AN....183..345V,1962P&SS....9..719L,1962AJ.....67..591K}
may induce high eccentricities in black hole binaries in AGN disks \citep[e.g.][]{2018ApJ...863...68L,2020ApJ...901..125D,2020ApJ...898...25T}. This effect arises due to the equal precession rates of $\Omega_b$ and $\varpi_b$ in Newtonian systems. For circular orbits with extreme mass ratios ($M_\bullet \gg m_b$), $\cos{i_b}\sqrt{1-e_b^2}$ is effectively conserved, and binary inclination may be exchanged for eccentricity when $|\cos(i_b)|<\sqrt{3/5}$ \citep{1962AJ.....67..591K}.
Within this work we restrict ourselves to binaries with $|\cos(i_b)|>\sqrt{3/5}$, deferring binaries which undergo ZLK cycles to subsequent studies.

\section{Numerical Methods}\label{sec:methods}
\begin{figure}
\includegraphics[width=\columnwidth]{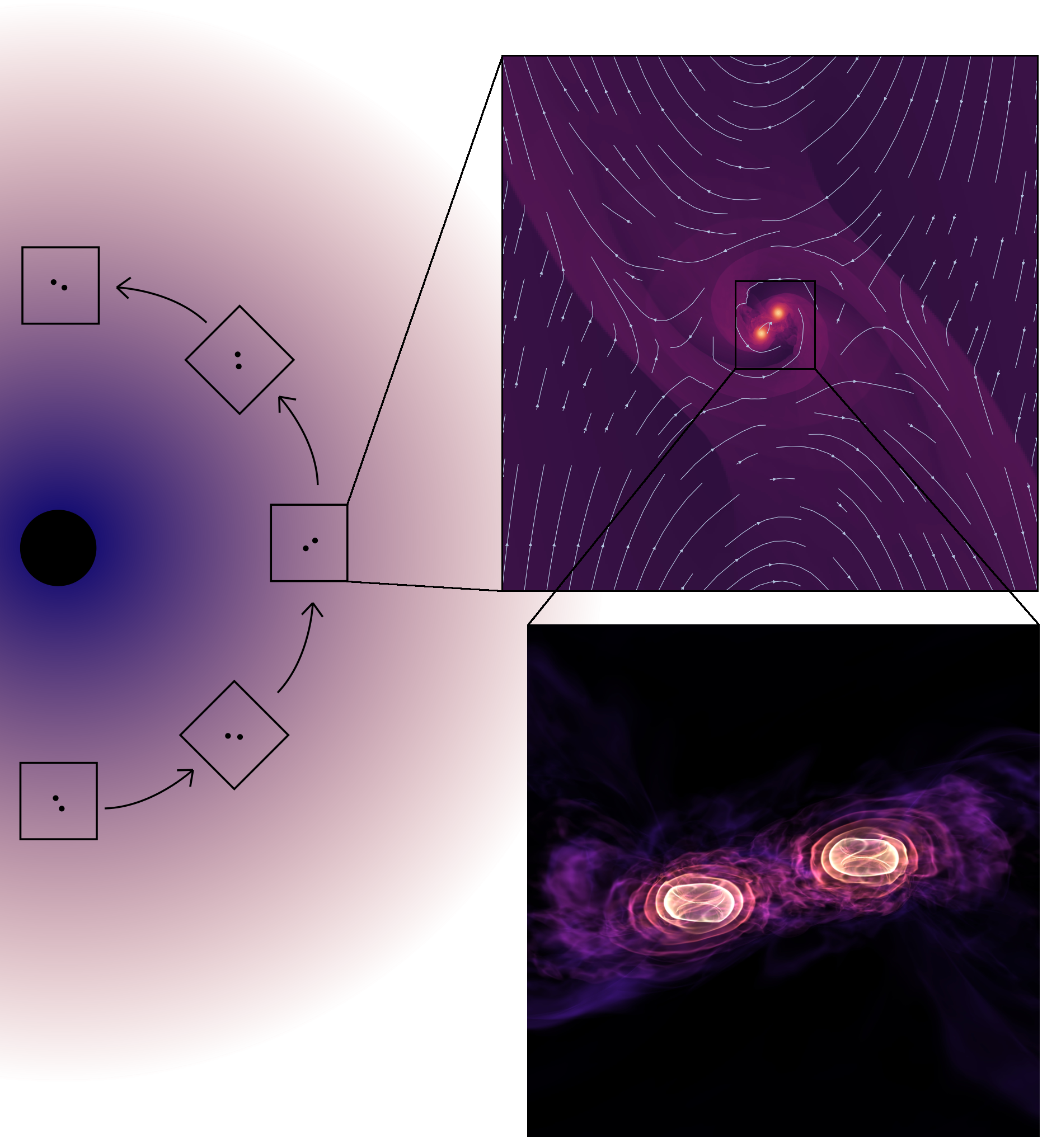}
\caption{A schematic diagram of our simulation methodology. We do not simulate the global AGN disk (left), but instead consider a shearing box which orbits the SMBH along with the binary. The upper right panel plots a density slice through the midplane during a simulation of an inclination $i=30^\circ$ binary, illustrating the quasi-global features captured by our simulation such as the excitation of spiral arms in the AGN disk and the horseshoe orbits of fluid elements which remain unbound from the binary. The bottom panel depicts opaque  density iso-contours in the immediate vicinity of the binary in the same simulation, demonstrating that the flow of gas in three dimensions is resolved well by our methodology. In both plots of simulation outputs, brighter colors indicate higher densities, which are colored on a logarithmic scale.}\label{fig:schematic}
\end{figure}

We simulate binaries embedded in AGN disks using the \texttt{Athena++} code \citep{2020ApJS..249....4S}, generally following the procedure described in \citet{2022ApJ...940..155D}. We assume that the gas is isothermal and inviscid, and use the shearing-box approximation \citep{1995ApJ...440..742H,1996ApJ...463..656S}, which expands the equations of hydrodynamics about a reference point in the disk, in our case the BHB center of mass. 
We illustrate this methodology schematically in Figure \ref{fig:schematic}, which highlights both the quasi-global features captured by our simulations, such as the horseshoe orbits followed by certain fluid elements and the spiral arms excited in the AGN disk, and the high fidelity with which the flow of gas around the binary is captured in three dimensions. 

The equations of compressible, isothermal hydrodynamics in the shearing box frame, along with source terms describing the interaction between the fluid and the BHB, are
\begin{equation}\label{eq:mass}
\frac{\partial\rho}{\partial t} + \nabla\cdot(\rho\mathbf{v}) = S_\Sigma,
\end{equation}
\begin{equation}\label{eq:momentum}
\begin{split}
\frac{\partial\mathbf{v}}{\partial t} + \mathbf{v}\cdot\nabla\mathbf{v} = -\frac{\nabla P}{\rho}-\nabla\Phi+\frac{\mathbf{S}_p}{\rho}\\-2\Omega_0\hat{\mathbf{z}}\times\mathbf{v}+3\Omega_0^2x\hat{\mathbf{x}}-\Omega_0^2z\hat{\mathbf{z}},
\end{split}
\end{equation}
where $\Phi$ is the gravitational potential of the BHB, $S_\Sigma$ is a mass sink term, and $\mathbf{S}_p$ is a momentum sink term. 
Each sink term $S_\Sigma=\sum_iS_{\Sigma,i},~\mathbf{S}_p=\sum_iS_{p,i}$ is localized to the vicinity of the BHB point masses.  
As illustrated in Figure \ref{fig:schematic}, we place the shearing-box at some orbital radius, $R_0$, in the AGN disk.
The coordinate basis vectors in this frame rotate in time and are $\hat{\mathbf{x}}$, which points from the SMBH to the center of the shearing-box; $\hat{\mathbf{z}}$, the direction normal to AGN disk midplane; and $\hat{\mathbf{y}}$ which is defined as $\hat{\mathbf{y}}=\hat{\mathbf{z}}\times\hat{\mathbf{x}}$ to complete the orthonormal triad. 
The rotation rate of the box is set to the Keplerian value, $\Omega_0 = \sqrt{GM_\bullet/R_0^3}$, and the binary center of mass is initially placed at the shearing box center, but we note that the binary center of mass is not fixed to this location.
In the remainder of the paper we use $\Omega_0$ and $n_\bullet$ interchangeably since $n_\bullet$ does not change appreciably over the course of a simulation. 
We close the equations of motion with an isothermal equation of state such that $P=c_s^2 \rho$, where the sound-speed is $c_s=H_0\Omega_0$, and $H_0$ is a scale height which we take to be constant throughout the domain.
The $\Omega_0^2z\hat{\mathbf{z}}$ term represents the vertical gravitational acceleration in the thin disk due to the SMBH. 
As discussed in Section \ref{sec:basic}, we neglect both the self-gravity of the disk and the force of the disk on the black holes.

We model the gravitational potential of each black hole using a spline function which is exactly Keplerian outside of a softening radius $r_s$, within which the potential is softened to avoid singularities \citep{2001NewA....6...79S}. We set the softening length to $r_s=0.04a_b$.\footnote{The spline gravitational softening described in \citep{2001NewA....6...79S} is typically employed using a softening length $\sim2.8$ times larger than a Plummer-softened potential because of the relative depths to which the potentials reach. The value used here is factor of three smaller than in \citet{2022ApJ...940..155D}, which we found to have appreciable consequences in our simulations of retrograde binaries.} We apply torque-free sink terms $S_\Sigma$ and $\mathbf{S}_p$ \citep[][]{2020ApJ...892L..29D,2021ApJ...921...71D} for gas within $r_s$ from either point mass as described in Appendix A of \citet{2022ApJ...940..155D}. 
We gradually introduce the black hole binary to the system over a timescale of $0.5n_\bullet^{-1}$, only allowing accretion after the conclusion of that period. We orient the binary as described in Section \ref{sec:def}, using the simulation $x$-axis as a reference direction and $x-y$ plane as a reference plane for defining $\Omega_b,$ $i_b$, and $\omega_b$. 

Our simulation domain extends from $-24H$ to $24H$ in $x$ and $y$, and from $-4H$ to $4H$ in $z$, resolving the scale height with five cells in each direction. Additionally, we utilize seven levels of static mesh refinement, allowing \texttt{Athena++} to automatically place additional refinement regions in order to maintain that neighboring cells differ by at most one refinement level. Accordingly, we resolve the gas around the binary with more than $1200$ cells per $a_b$. Our simulations used 2nd-order spatial reconstruction \citep{1974JCoPh..14..361V}, the HLLE approximate Riemann solver \citep{10.2307/2030019,1988SJNA...25..294E}, and a second-order strong stability preserving Runge-Kutta time integrator \citep[RK2][]{1998MaCom..67...73G}. 
We found that the Runge-Kutta integrator leads to a much more stable time step than the 2nd-order van Leer integrator \citep{2009NewA...14..139S} used in \citet{2022ApJ...940..155D}, enabling us to use much higher resolution and smaller softening lengths.\footnote{With these savings, each standard simulation only used approximately 400,000 core hours on the Chicoma supercomputer at Los Alamos National Laboratory.}

Our simulations were initialized according to the background equilibrium solution, with $v_y=-3\Omega_0x/2$, $v_x=v_z=0$, and $\rho=\rho_0\exp{[-(z/H)^2/2]}$.
Because we neglect the self-gravity of the disk and disk-induced binary motion, our simulations are scale-free and the value of $\rho_0$ is set to unity without loss of generality.

We used shear-periodic boundary conditions in the $x-$direction and periodic boundary conditions in the $y-$direction. In the $z-$direction we applied standard outflow boundary conditions to the velocities, but extrapolated the density so that vertical hydrostatic equilibrium could be maintained. 

\begin{figure}
\centering
\includegraphics[width=\linewidth]{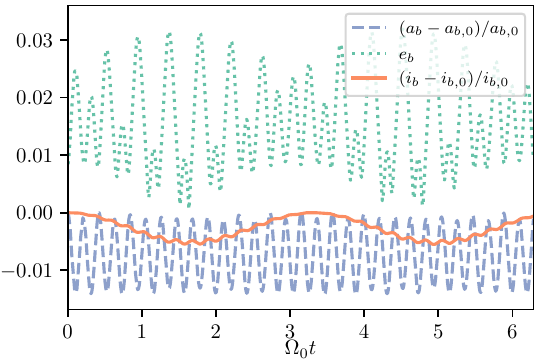}
\caption{Evolution of the semi-major axis (blue dashed), inclination (orange solid), and eccentricity (green dotted) of our $i_b=165^\circ$ binary over the course of one orbit about the SMBH.}
\label{fig:orbit165}
\end{figure}

\subsection{N-body evolution}
Previous studies of embedded BHBs either use fixed binary orbits with ad hoc precession terms \citep[e.g.][]{2022MNRAS.517.1602L} or solved the 3-body equations of motion under simplifying assumptions \citep{2022ApJ...940..155D,2023arXiv230911561W}. 
We found that the approximate N-body evolution scheme employed in \citet{2022ApJ...940..155D}, while sufficient for circular $i_b=0$ binaries, did not properly conserve angular momentum for inclined binaries and resulted in ZKL cycles of erroneous magnitude, motivating the present treatment.
The difficulty lies in that fact that even for mildly hierarchical triples (e.g., $a_b\sim R_H/4$) such as the ones presented in Section \ref{sec:results}, the orbital evolution of the binary is non-trival. 
We illustrate this by the evolution of the orbital elements of an $i_b=165^\circ$ binary in Figure \ref{fig:orbit165}.

To ensure that we evolve the BHB accurately, we have developed an N-body extension to  \texttt{Athena++} that fully couples the evolution and interaction of N gravitating and accreting bodies to the hydrodynamics solver. 
Integration of the gravitational system is done with a full \texttt{REBOUND} \citep{2012A&A...537A.128R} simulation embedded in \texttt{Athena++} and uses the IAS15 integrator, an implementation of a 15th-order Radau quadrature \citep{Radau1880,2015MNRAS.446.1424R}.
During one \texttt{Athena++} time step, \texttt{REBOUND} evolves the positions and velocities of all objects in the simulation to the appropriate end time of each stage of the integrator. 
For example, for the RK2 integrator used here, the two-stage update equations for the schematic system $dU/dt = F(U)$ are

\be
 U^{(1)} &=& U^{(0)} + \Delta t F(U^{(0)})  \\
 U^{(2)} &=&  \frac{1}{2} \left( U^{(0)} +  U^{(1)} \right) + \frac{\Delta t}{2} F(U^{(1)}) 
\ee
such that the final updated values are,
\be
 U^{(2)} = U^{(0)} + \frac{\Delta t}{2} \left( F(U^{(0)}) + F(U^{(1)}) \right)
\ee
and where $U^{(0)},U^{(1)},$ and $,^{(2)}$ are the values of $U$ at the beginning of the step, at the end of the first stage, and at the end of the second stage, respectively. For our simulations $U$ is the vector of conserved variables, as well as the position and velocities of all gravitating bodies in the \texttt{REBOUND} simulation.

During stage 1, we first measure the force between each body and the gas for the positions and velocities at the beginning of the step (i.e., evaluating $F(U^{(0)}$) and then evolve the N-body system to the end of the step. 
During stage 2, we measure the force between each BH and the gas for the positions and velocities at from the end of stage 1 (i.e, evaluating $F(U^{(1)})$).  
For each stage we add the force on each BH to a running time average with the appropriate weights.
With this method we are able to achieve 2nd order in time coupling between rebound and \texttt{Athena++} for the RK2 integrator. 
We have also tested our method on the RK1, VL2, and RK3 time integrators available in \texttt{Athena++}.
The \texttt{REBOUND} simulation is always performed in the “global” frame of the AGN disk, but the \texttt{Athena++} simulation is performed in the “local” shearing-box frame. 
We thus need to rotate the \texttt{REBOUND} results into the local frame when computing the gas forces. 

With \texttt{REBOUND} taking care of the numerical integration of the N-body system between each hydrodynamical time step we are able to accurately model many systems of interest for embedded Black holes in AGN disks that were previously too numerically difficult to evolve. Examples include highly eccentric and inclined binaries, small clusters of objects, arbitrarily complicated inner and outer binary configurations, and binary formation. 
Our sub-stage coupling method also allows for accurate book-keeping of the forces on the BHs allowing for high-fidelity time averages of orbital evolution quantities.

\section{Results}\label{sec:results}
\begin{figure}
\centering
\includegraphics[width=\linewidth]{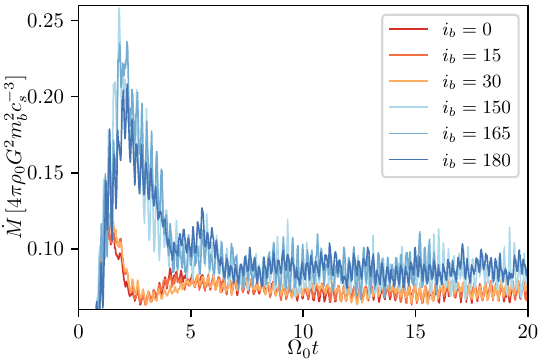}
\caption{The accretion rate onto each binary, normalized to the Bondi accretion rate $\dot{M}_B=4\pi\rho_0G^2m_b^2/c_s^3$ \citep{1952MNRAS.112..195B}, over the course of our simulations. Retrograde ($i_b>90^\circ$) binaries accrete at higher rates than prograde ($i_b<90^\circ$) binaries. We use shades of red and orange to plot the accretion rate for prograde binaries, while we plot the accretion rate for retrograde binaries using shades of blue, in each case using lighter shades for binaries closer to $90^\circ$.}
\label{fig:mdots}
\end{figure}

We present six simulations over a range of binary inclinations, $i_b\in\{0,15^\circ,30^\circ,150^\circ,165^\circ,180^\circ\}$. 
We focus on a single binary configuration, with mass ratio $m_b/M_\bullet=1.536\times10^{-6},$ separation $a_b=R_H/4$ and a disk scale height of $H=0.01 R_\bullet$ such that $R_H/H=0.8$ \citep[c.f.][]{2022ApJ...940..155D}. 
In each case, the mean motion of the binary was $n_b \sim 13.9 n_\bullet$.
Each binary was initialized with $e_b=0$, $\omega_b=0$, and $\Omega_b=0$.
The simulations were run for at least $20n_\bullet^{-1}$, or $\sim3$ orbits of the binary around the SMBH, which was sufficient for the accretion rates onto the binary to reach a quasi-steady state, as illustrated by Figure \ref{fig:mdots}. 
We perform all of our measurements of orbital evolution at times later than $14n_\bullet^{-1}$, well after initial transients have died out. 

We focus first on somewhat large scales ($\sim R_H$) in Section \ref{sec:large}, at which the influence of the binary begins causing deviations in the flow compared to a single black hole of the same mass. We investigate the flow of gas through the accretion torus (minidisk or circum-single disk, `CSD') surrounding each black hole in Section \ref{sec:minidisks}, and binary orbital evolution in Section \ref{sec:orbits}.

\begin{figure*}
\includegraphics[width=\linewidth]{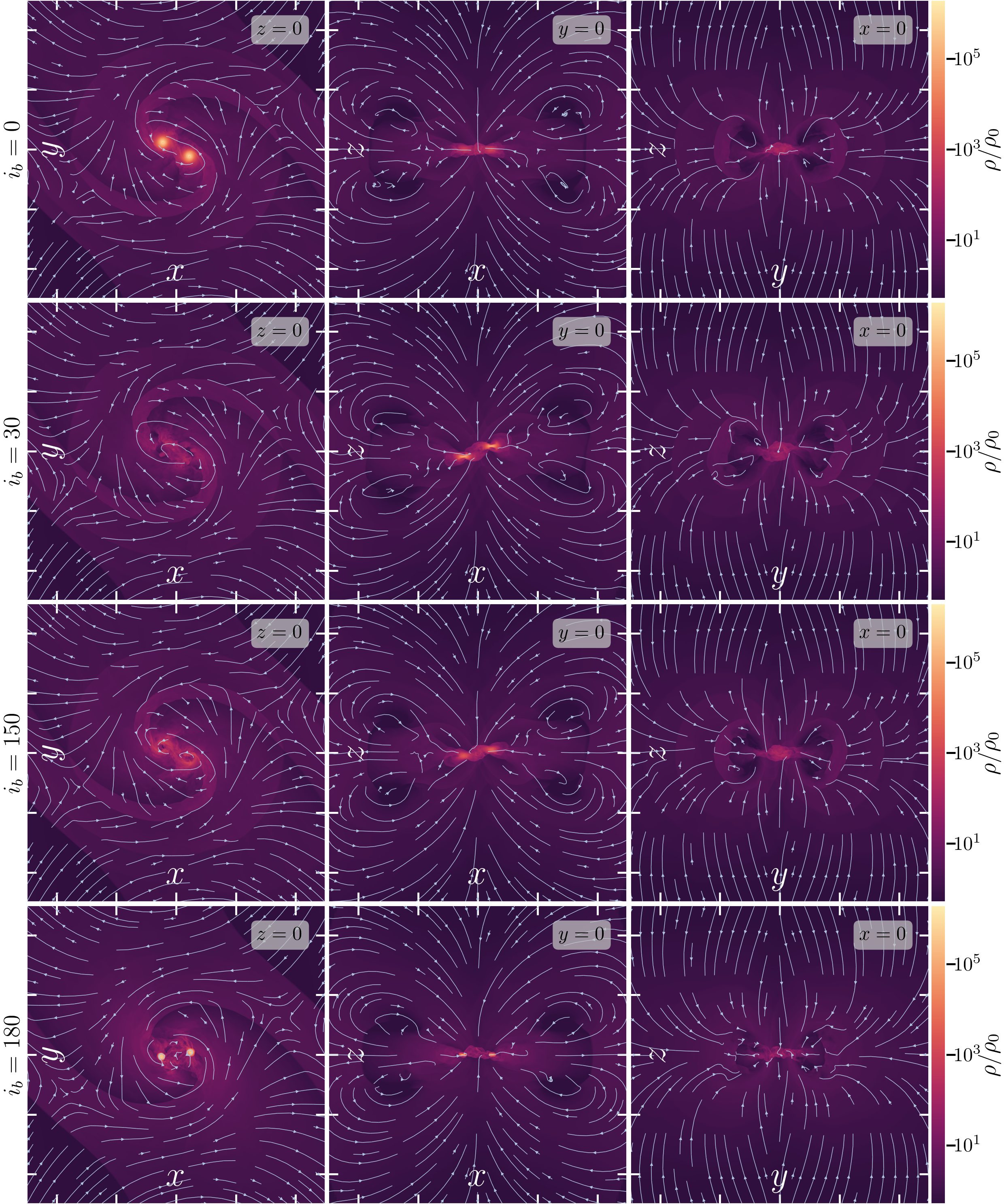}
\caption{Slices of density along each coordinate plane, along with velocity streamlines, at $t=14.5 n_\bullet^{-1}$ in our 
$i_b=\{0^\circ,30^\circ,150^\circ,180^\circ\}$ simulations. The first, second, and third columns plot data in the $x-y$, $x-z$, and $y-z$ planes respectively. 
Curved light blue arrows illustrate streamlines of the in-plane velocity field in each slice. In all cases, it is evident that accretion along the vertical direction plays a significant role. Additionally, there is evidence for an $m=2$ spiral in each case, although these spirals are weaker when $i_b>0^\circ$, and the deflection of fluid elements on approximate horseshoe trajectories is clearly visible through the streamlines in the $x-y$ plane. The axis ticks are spaced $2a_b=R_H/2$ apart, and each panel roughly encompasses a slice through the Hill sphere of each binary.}
\label{fig:slices}
\end{figure*}

\subsection{Large-Scale Structures}\label{sec:large}
On scales much larger than the binary semi-major axis, the presence of the binary has a largely negligible affect on the flow compared to a single black hole of the same mass. For example, the fluid density slices in Figures \ref{fig:schematic} and \ref{fig:slices} demonstrate the formation of a spiral arm in the larger disk. The streamlines in the aforementioned figures illustrate the horseshoe orbits followed by fluid elements that are deflected away from the binary, and the separatrices in the midplane velocity field where fluid is captured by the binary. 

However, because the binary separation is only a quarter the size of the Hill radius, the binary has a non-negligible influence on the gas within the Hill sphere, as illustrated for binaries at four different inclinations in Figure \ref{fig:slices}.
Turning first to the density distribution in the disk midplane, it is clear that the binary creates $m=2$ spirals at larger scales - while for prograde binaries these extend down and into the CSDs, these spirals are not as well defined within $r\sim a_b$ about retrograde binaries. Contrasting the simulations of aligned ($i_b=0^\circ$) and anti-aligned ($i_b=180^\circ$) binaries, while the gas just within the Hill sphere is prograde in both cases, the gas in the minidisks of the aligned binary orbits in a prograde sense while the minidisks of the anti-aligned binary orbit in a retrograde sense. The minidisks around retrograde binaries are significantly smaller than those of their prograde counterparts, largely due to the intense ram pressure experienced by gas orbiting along with the binary against gas falling in from larger scales in a prograde sense, which we investigate further in Section \ref{sec:minidisks}.

Turning to the slices through the $x-z$ and $y-z$ planes displayed in the second and third columns of Figure \ref{fig:slices}, we observe large-scale circulation, especially along the axis connecting the binary and SMBH. Although the precise dynamics depend on the binary inclination, matter more than $\sim2a_b$ from the binary tends to be pushed away from the binary and subsequently lifted upwards to $|z|\gtrsim a_b$, joining the meridionally circulating flow which accretes onto the binary from higher latitudes. Much of the accretion onto the binary occurs along the vertical direction, necessitating three-dimensional study of these accretion flows. The circulation pattern observed here is qualitatively similar to that observed in some three-dimensional simulations of planets embedded in circumstellar disks \citep[e.g.][]{2014ApJ...782...65S,2016ApJ...832..105F}, as well as the previous simulations of aligned embedded binaries \citep{2022ApJ...940..155D}. 

One peculiar feature of retrograde binaries is their higher accretion rates relative to their prograde counterparts, as illustrated in Figure \ref{fig:mdots}: prograde binaries typically accrete at about 7.2\% of the Bondi rate, whereas retrograde binaries typically accrete at about 8.6\% of the Bondi rate. A similar trend was noted in \citep{2022MNRAS.517.1602L}, which attributed the increased accretion rate onto retrograde binaries to those binaries lacking CSDs, suggesting that the presence of minidisks around prograde binaries decreases the accretion rate. Unlike \citep{2022MNRAS.517.1602L}, we observe CSDs around retrograde binaries,\footnote{However, the CSD \emph{size} in our simulations of retrograde is not converged. We illustrate this in Appendix \ref{sec:appendixSoft}, and show that this does not affect any of our other results.} so their existence or lack thereof cannot dictate these trends in accretion rate. Moreover, our $i_b=150$ binary has both the smallest minidisks and \emph{lowest} accretion rate out of the retrograde binaries studied here.

We have found that throughout the Hill sphere of each binary, to distances $\gtrsim R_H=4a_b$, the average fluid density is lower and average radial velocity towards the binary barycenter is higher around retrograde binaries. Because gas typically enters the Hill sphere orbiting in a prograde sense, the prograde binaries will have, on average, longer-duration gravitational interactions with gas, potentially doing more work and hindering the binding of gas to the binary. Regardless, the higher accretion rates onto retrograde binaries begin at large distances from the binary and do not depend on the size of the minidisks.

\subsection{Minidisks}\label{sec:minidisks}

\begin{figure}
\includegraphics[width=\linewidth]{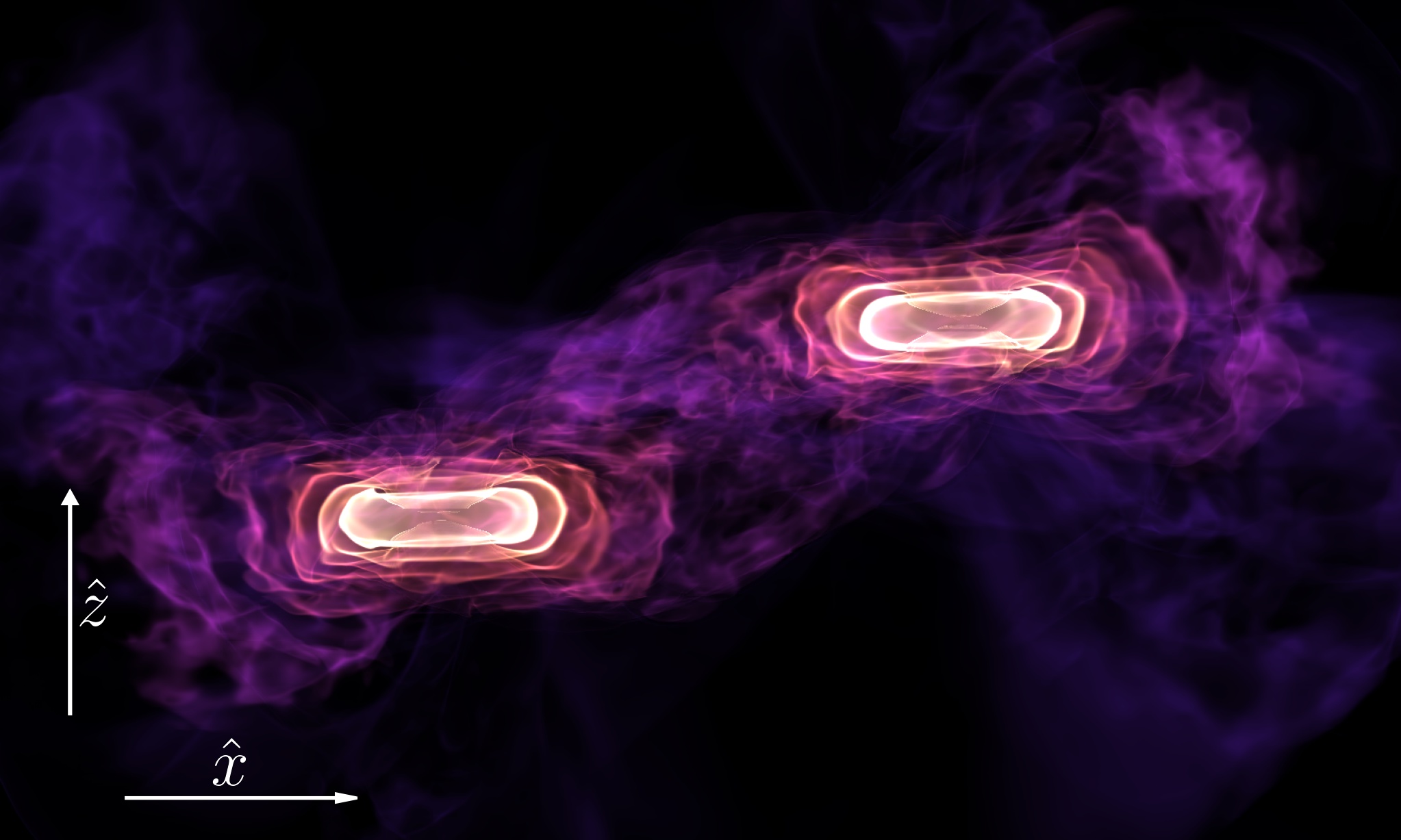}
\caption{A volume rendering of the gas around the black holes in our $i_b=30^\circ$ simulation at $t=14.5n_\bullet^{-1}$. The darkest colors highlight gas at densities of $\rho\sim\rho_0/10$, and the brightest colors highlight gas at densities of $\rho\sim10^4\rho_0$, with intermediate colors spaced log-uniformly in density. Even though the orbit of the binary is inclined relative to the AGN disk midplane, the minidisks remain oriented along with the AGN disk rather than the plane of the binary orbit. Notably, the lower-density gas flowing away from the binary is in closer alignment with the binary orbital plane.}
\label{fig:csd30}
\end{figure}

The accretion disks that form around each black hole in our simulations display a number of notable characteristics: these mindisks exhibit meridional circulation, gas outflowing along the midplane and inflowing at higher latitudes, similar the flows on the much larger scales of the Hill sphere shown in in Figure \ref{fig:slices}. 
The accretion disks that form around the black holes in prograde binaries orbit in a prograde sense, whereas the accretion disks that form around retrograde binaries form in a retrograde sense (that is, with the same handedness of the binary: $\mathbf{l}_b\cdot\mathbf{l}_{\rm gas}>0$); and the minidisks themselves are typically aligned with  the AGN disk midplane - that is to say the angular momentum vector of gas in a minidisk around a prograde binary is nearly parallel to $\vec{z}$, and the angular momentum vector of gas comprising the CSD of a retrograde binary is nearly parallel to $-\vec{z}$. Particularly for inclined binaries, these CSD alignments could appreciably alter the effective spin distribution of AGN-channel black hole binary mergers \citep[e.g][]{2023arXiv230915213M}.  As an example, a volume rendering of the minidisks of a $i_b=30^\circ$ binary are displayed in Figure \ref{fig:csd30}. Notably, while the gas most proximate to each black hole is aligned with the AGN disk midplane, the lower-density gas at larger distances, corresponding to the spirals and outflows seen in Figure \ref{fig:slices}, is more aligned with the orbital plane of the binary. We caution that this result does not hold at all binary inclinations: preliminary simulations of more misaligned binaries, which we plan to present in a subsequent paper, suggest that the minidisks of those binaries will align more with the binary orbital plane than do the minidisks of binaries in the present investigation.

\begin{figure*}
\includegraphics[width=\linewidth]{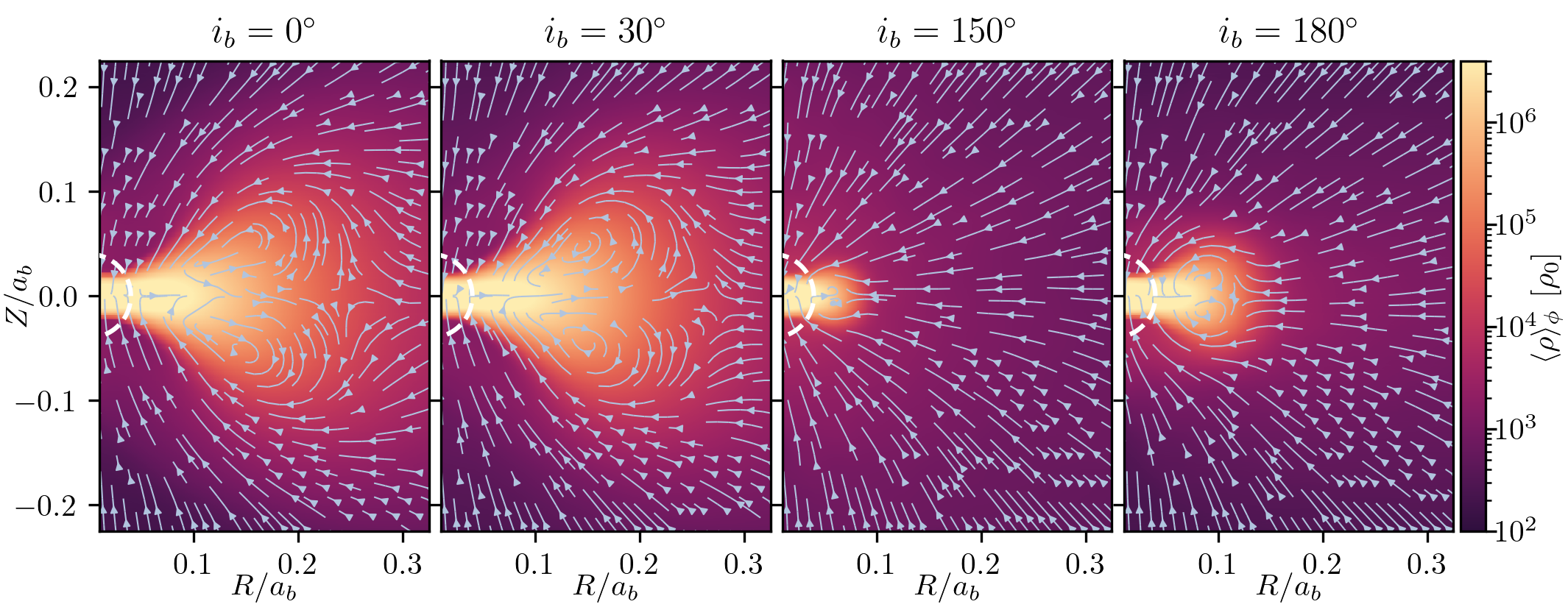}
\caption{Profiles, averaged in time and azimuth, of the gas density and velocity in a cylindrical polar coordinate system in the frame of one of the black holes from each of our $i_b\in\{0^\circ, 30^\circ, 150^\circ, 180^\circ\}$ simulations. Streamlines illustrate meridional circulation in the minidisks and vertical accretion onto the binary through the direction of the azimuthally-averaged velocity field, and color indicates the azimuthally-averaged fluid density. Time-averaging used 27 individual snapshots over the course of three binary orbits. The dashed white circle denotes the region where our sink particle operates and the gravitational potential is softened.}
\label{fig:rhostream}
\end{figure*}

\begin{figure*}
\includegraphics[width=\linewidth]{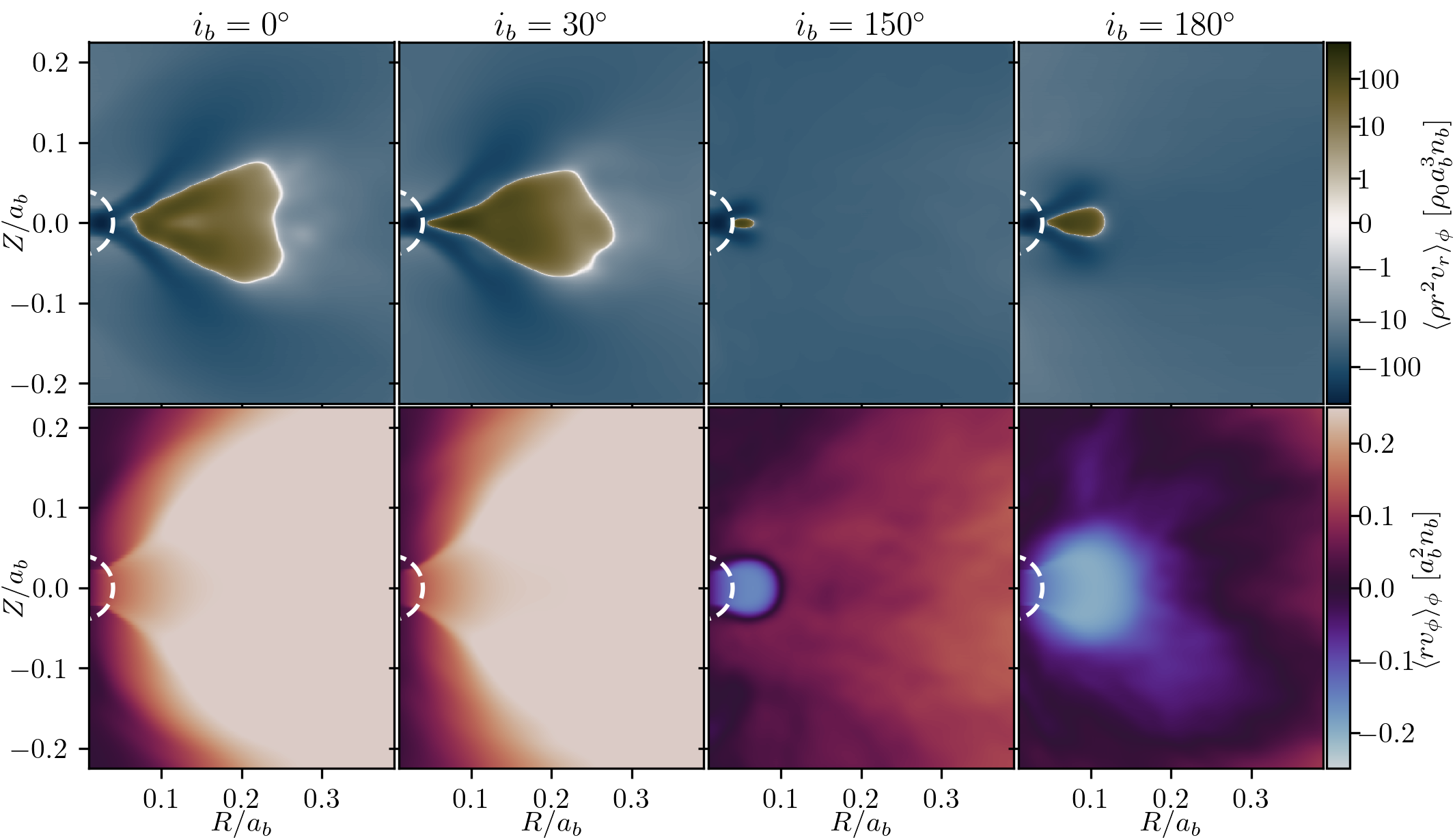}
\caption{Azimuthally averaged profiles of radial mass flux (top row) and the specific angular momentum of the gas (bottom row) in a cylindrical polar coordinate system in the frame of one of the black holes from each of our $i_b\in\{0^\circ, 30^\circ, 150^\circ, 180^\circ\}$ simulations. Time-averaging used 27 individual snapshots over the course of three binary orbits. The dashed white circle denotes the region where our sink particle operates and the gravitational potential is softened. }
\label{fig:velstuff}
\end{figure*}

It is worth comparing the alignment of these minidisks to the disk-binary alignment observed in earlier simulations of viscous, isolated circumbinary disks \citep[e.g.][]{2019ApJ...875...66M}: in such simulations the binary accretes from a circumbinary disk much thinner than the orbit of the binary, and (in a frame oriented with the circumbinary disk) out-of-plane velocities in the circumbinary disk are negligible compared to the orbital motion of the binary. However, embedded binaries accrete gas vertically, and over the course of a binary orbital period just as much matter falls on from above as below. Gas flows onto isolated binaries from their circumbinary disks, whereas gas falls onto embedded binaries ballistically \citep{2022ApJ...940..155D}. Additionally, \citet{2023arXiv230518538A} recently identified minidisk tilt oscillations on the order of the disk aspect ratio. 

The gas constituting the minidisks of embedded binaries is also not particularly long-lived. While the typical viscous inflow timescale of material in the minidisks of isolated binaries can be tens to hundreds of binary orbital periods,\footnote{A simplistic estimate follows from treating each CSD as a circular Keplerian disk with a radius matching the size of the Roche lobe of each black hole, in which case $r/v_r\approx91n_\bullet^{-1}\approx15t_{\rm orb}$ for an unrealistically high viscosity of $\nu=10^{-3}a_b^2\Omega_b$ and equal-mass binary.} we find that the mass-weighted average inflow timescales within the Roche lobe of each black hole are on the order of $\sim0.2-4$ binary orbital periods $t_{\rm orb}$ in the minidisks themselves and $\lesssim 0.1t_{\rm orb}$ in the polar regions around each black hole. We note, however, that for gas to form a CSD around a black hole, it must necessarily be moving with a similar bulk velocity to that black hole, lest it be left behind or captured by the other black hole; thus the gas forming each CSD has significantly torqued the black hole around which it orbits, even if the orbital plane of that gas does not align with the binary orbital plane. 

The flow of gas onto and throughout the minidisks over a range of inclinations is illustrated in Figure \ref{fig:rhostream}, which plots the azimuthally averaged gas density (in the frame of one of the black holes) along with velocity streamlines in the same frame. Although quasiballistic accretion occurs onto prograde binaries most prominently at higher latitudes, such accretion occurs over a wider range of angles in retrograde binaries until reaching the minidisk. Intriguingly, the CSDs surrounding the black holes comprising the $i_b=150^\circ$ binary are visibly smaller than those around the $i_b=180^\circ$ binary, which are in turn smaller than those of the black holes in prograde binaries. In fact, the CSD of the $i_b=150^\circ$ binary is only about twice as large as the size of our sinks. However, the CSDs themselves display meridional circulation, with high-latitude inflows and midplane outflows, regardless of their size. 

Another view of the flow of gas throughout and around the minidisks is provided in Figure \ref{fig:velstuff}, which plots the time- and azimuth-averaged profiles of the gas specific angular momentum and radial mass flux. Patterns of meridional circulation and midplane outflows are again evident in the distribution of radial mass flux, which also highlights accretion through the polar regions. Even though the CSDs around the black holes in the $i_b=150^\circ$ binary are very small ($\lesssim a_b/10$ in extent), they still display the same circulation patterns. Although the specific angular momentum profile is not particularly informative for prograde binaries, around the black hole in retrograde binaries it very clearly demarcates which gas is flowing in a retrograde sense (and thus with the same handedness as the orbit of the binary) compared to gas orbiting in a prograde sense. 

\begin{figure}
\includegraphics[width=\linewidth]{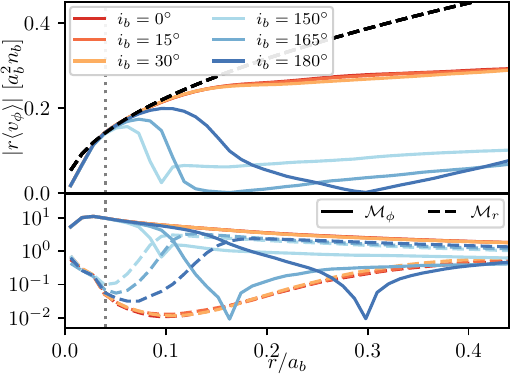}
\caption{One-dimensional profiles, averaged over spherical shells centered on each black hole, of the fluid specific angular momentum (top panel) and both radial and azimuthal fluid Mach numbers (bottom panel). Warm colors display results for prograde binaries and cool colors display results for retrograde binaries. In the top panel, the dashed black like displays the specific angular momentum profile of a Keplerian disk. In the bottom panel the radial Mach number is plotted using dashed lines and the azimuthal Mach number is plotted using solid lines. We have calculated the characteristic Mach number at a given radius by integrating the radial and azimuthal momentum within each shell, dividing by mass in that shell and the sound speed, and taking the absolute value of the result. The vertical dotted line indicates the radius within which we apply sink terms and gravitational softening.}
\label{fig:profs1d}
\end{figure}

A slightly more condensed view of the velocity profile throughout the Roche lobe of each black hole is shown in Figure \ref{fig:profs1d}. The top panel therein illustrates the specific angular momentum profile in comparison to a reference Keplerian curve. Prograde binaries have sub-Keplerian disks at $r\gtrsim a_b/10$, and the time-averaged flow does not depend substantially on the binary inclination. The bottom panel illustrates the azimuthal and radial Mach numbers ($\mathcal{M}_\phi$ and $\mathcal{M}_r$ respectively) of the gas around each black hole. As characteristic of a disk, the azimuthal flow is supersonic throughout, varying as $\mathcal{M}_\phi\propto r^{1/2}$, and in the prograde disks the radial Mach number is everywhere both below unity and orders of magnitude below the azimuthal Mach number. The Mach number profiles of CSDs around prograde binaries show very little variation with inclination. 

\begin{figure}
\includegraphics[width=\linewidth]{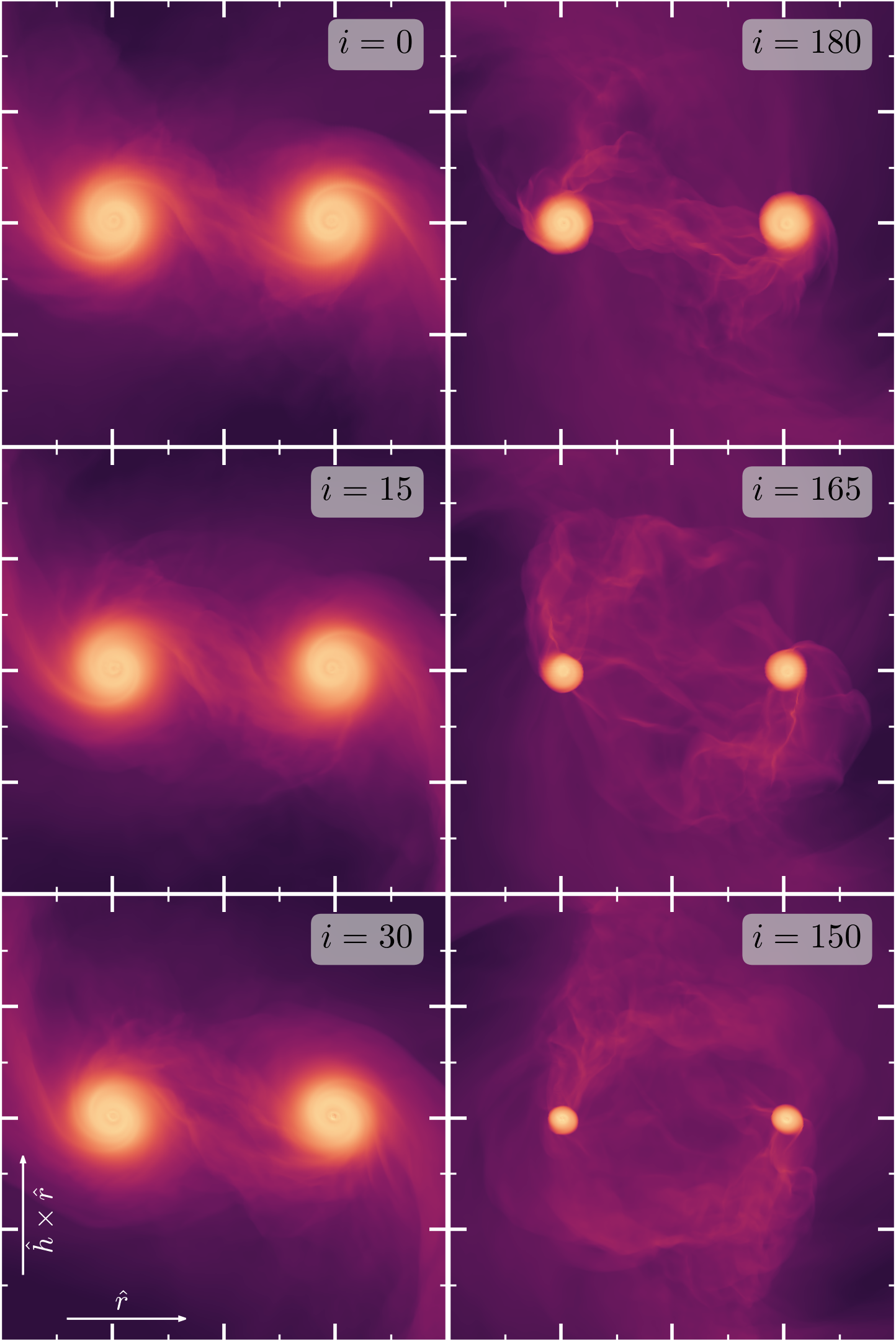}
\caption{Projections of the fluid density along the orbital angular momentum vector of each binary at $t=18.4n_\bullet^{-1}.$; in each case the angular momentum vector of the binary is pointing out of the page, and in this frame the binary is moving in a counterclockwise sense. Because the CSDs are aligned with the AGN disk midplane rather than the orbital plane of the binary, the minidisks of inclined binaries appear slightly prolate in these projections. We can clearly see the smaller minidisks around the retrograde binaries, as well as the streams of gas being stripped off of the retrograde binaries. Major ticks are placed $0.5a_b$ apart, with minor ticks every $0.25a_b$.}
\label{fig:projecting}
\end{figure}

The flow of gas around the black holes in retrograde binaries, however, is markedly different. The top panel of Figure \ref{fig:profs1d} illustrates how the size of the CSD around each black hole varies with binary inclination, smallest for the $i_b=150^\circ$ binary and largest for the $i_b=180^\circ$ binary, and how the disks are still Keplerian at small radii despite their smaller size. The bottom panel shows how fluid motion is generally azimuthally subsonic prior to forming a disk, but plunges in a radially supersonic manner up until that point. The gas initially falls in ballistically, with $\mathcal{M}_r\propto r^{-1/2}$, before forming an accretion disk.

The flow of gas onto, off of, and between the minidisks in each binary is displayed in Figure \ref{fig:projecting}, which plots a projection of the fluid density along the binary angular momentum vector. The CSDs around prograde binaries appear very similar to those observed in simulations of standard circumbinary disks \citep[e.g.][]{2017ApJ...835..199R,2019ApJ...871...84M,2021ApJ...921...71D}: spiral waves are present throughout the disks, some matter flows onto the binary from larger distances, and streams of gas flow between the two black holes. Although some very faint spirals can be seen in the CSDs of retrograde binaries, they are far less prominent \citep[cf.][]{2021ApJ...911..124L}. As presaged by Figures \ref{fig:velstuff} and \ref{fig:profs1d}, ample gas appears to be stripped away from the retrograde binaries, contributing to their smaller sizes; this trailing gas appears to be more substantial for the binaries that orbit outside of the AGN disk midplane, in line with the observed trends in CSD size. 

\subsection{Orbital evolution}\label{sec:orbits}
Based on our examinations of the flow of gas onto the binary and around the black holes constituting it, we can draw a number of qualitative inferences. As a first approximation, we can assume that gas accreting onto each black hole does so on average with the same velocity as that black hole. Thus, the accretion of gas onto the binary should add to its specific energy largely through changing the mass of the binary and thus its specific gravitational binding energy, and for circular binaries with $|\vec{r}|\approx a_b$, $\dot{\mathcal{E}}/\mathcal{E}\approx 2\dot{m}_b/m_b$. However, gas must journey deep into the potential well of an individual black hole before accreting, potentially loosing energy to the binary in the process and helping unbind it. For more tightly bound binaries, more work must be done on gas before it can accrete, and for sufficiently bound binaries this may cause binaries to outspiral as found in \citet{2022ApJ...940..155D}.

\begin{figure}
\includegraphics[width=\linewidth]{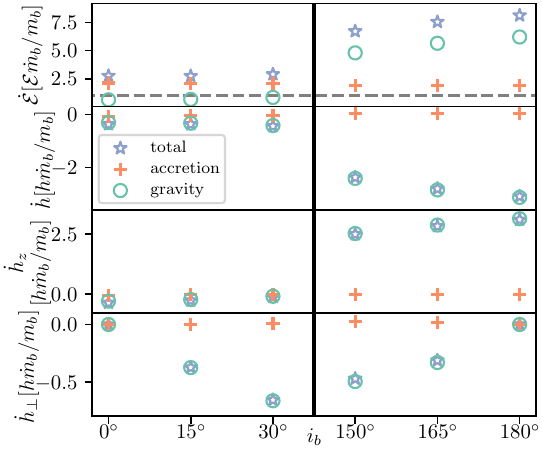}
\caption{The top row plots the rate of change of the binary specific energy normalized by the average binary specific energy and accretion timescale $m_b/\dot{m}_b$ for each of our simulations (the total rate shown by blue stars, the contribution due to accretion by orange crosses, and the contribution due to gravity by green circles). The second row plots the average rate of change of the magnitude of the binary specific angular momentum, the third the rate of change of the z-component, and the bottom row shows the rate of change of the perpendicular component, all normalized by the binary specific angular momentum and accretion timescale. In the first row, the gray dashed line indicates the critical value $\dot{\mathcal{E}}/\mathcal{E}=\dot{m}_b/m_b$ above which equal-mass binaries are driven to inspiral. All quantities were averaged in time between $14\,n_\bullet^{-1}$ and $20\,n_\bullet^{-1}$.}
\label{fig:physical}
\end{figure}

The above reasoning is directly applicable to prograde binaries, but must be adapted slightly for retrograde binaries. For example, if CSDs do not form at all \citep[e.g.][]{2022MNRAS.517.1602L}, then the assumption that gas accretes with the same velocity as a given black hole breaks down, and the kinetic contribution to the time derivative of the binary specific energy will become nonzero. On the other hand, if CSDs form around each black hole, as in our simulations, then the velocity of the gas must be significantly altered before it can accrete, as shown in Section \ref{sec:minidisks}, and thus the binary will do a significant amount of work on the gas. Similarly, as shown in Figure \ref{fig:velstuff}, the angular momentum of gas must change dramatically for it to become part of the minidisk, in which case the binary will experience a strong torque, greatly reducing the magnitude of its angular momentum, although it is not clear a priori if this will predominantly affect the binary eccentricity or semi-major axis. And similarly, for gas in the AGN disk which is on average aligned with the AGN disk midplane to accrete onto either black hole, that gas must gain significant out-of-plane angular momentum, and thus damp the inclination of the binary. 

With the aforementioned notions in mind, we turn to our measurements of the rate of change of the specific energy and angular momentum of each binary in our simulations in Figure \ref{fig:physical}. Turning first to the energy, our prior expectations are confirmed: in all cases the contribution to $\dot{\mathcal{E}}/\mathcal{E}$ due to accretion is $\sim2$, and while the contribution due to gravity is small in the prograde case, it is quite large for retrograde binaries, suggesting that the binaries should inspiral rapidly. Also in line with our qualitative expectations, accretion does not change the specific angular momentum of any binary.\footnote{Notably, if we used a naive sink prescription and allowed the sink particles to exert a torque on gas in the frame of each sink, then this would not be the case: Fluid angular momentum would instead be transferred to both the spin and orbit of the black holes \citep{2020ApJ...892L..29D,2021ApJ...921...71D}.} As argued above, retrograde binaries experience significant torques, especially contributing to changes in the $z-$component of their angular momenta. 

\begin{figure}
\includegraphics[width=\linewidth]{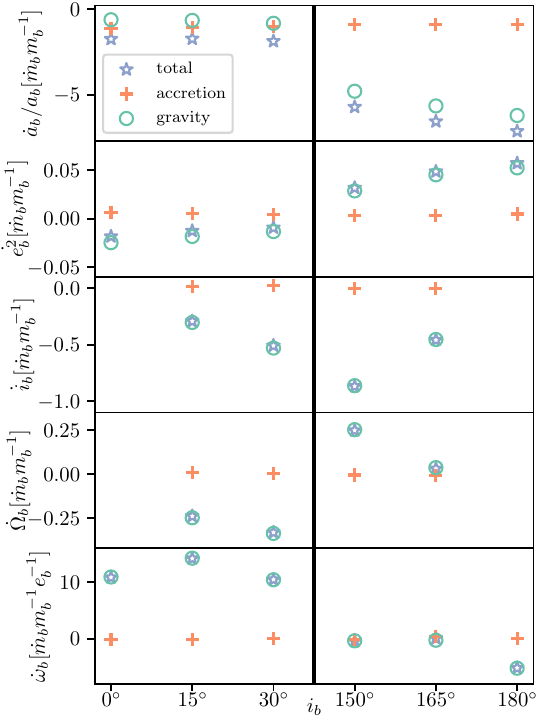}
\caption{The rates of change (due to gravity, accretion, and in total) of the binary semi-major axis, squared eccentricity, inclination, argument of the ascending node, and argument of periapsis. 
All quantities were averaged in time between $14\,n_\bullet^{-1}$ and $20\,n_\bullet^{-1}$. 
When reporting values of $\dot{i}_b$, $\dot{\omega}_b$, and $\dot{\Omega}_b$, we use radians rather than degrees. We only report the contributions to orbital elements due to accretion and gravitational interactions with the surrounding gas, which are independent of choices like the reference plane for defining $\Omega_b$.}
\label{fig:orbital}
\end{figure}

The orbital evolution of each binary follows from the above discussions. First, in the top row of Figure \ref{fig:orbital}, we see that although all binaries contract (at this separation, see \citet{2022ApJ...940..155D}), retrograde binaries do so at a much higher rate. Similarly to previous studies we find that eccentricity is excited in retrograde binaries \citep[e.g.][]{2015ApJ...806...88S,2023arXiv230703775T}, while eccentricity is damped in near-circular prograde binaries \citep{2019ApJ...871...84M,2021ApJ...909L..13Z}. Although $\dot{i}_b$ is not well-defined for $i_b=0$ and $i_b=180$, we see as expected that binary inclinations are damped. Otherwise put, this result suggests that any binary at all misaligned with the midplane of the AGN disk should be gradually aligned with the disk, in a prograde sense, over time. 
Thus, it appears that $i_b=0$ is a stable equilibrium while $i_b=180^\circ$ is unstable. If the trends we observe here continue to higher-inclination systems, initially near-retrograde binaries could realign with the AGN disk within a few mass-doubling timescales.

The time derivative of $\dot{\omega}_b$ is not particularly well-defined for near-circular binaries (due to the terms in the denominator of Equation (\ref{eq:omegadot}) that go to zero as $e_b\rightarrow0$), so we have normalized those results by the time-averaged values of $e_b$; for these effectively circular binaries, we simply emphasize that $\dot{\omega}_b$ is generally nonzero. The time derivative of $\Omega_b$ is more well defined, at least for binaries with $h_\perp\neq0$, and we find that it is negative for prograde binaries and positive for retrograde binaries. These signs relate to the handedness of the binaries passage through the disk midplane and the background shear of the AGN disk, and as we discuss in Section \ref{sec:discorb}, the important conclusion is that this disk-induced precession rate is nonzero. 

\section{Discussion}\label{sec:discussion}
In this section, we briefly discuss the implications of our results on the manner in which accretion onto embedded black holes proceeds, the dynamics of jets which may be formed during the accretion process, and the orbital evolution of embedded BHBs. We caution that the present work has eschewed magnetic fields, the effects of radiation, gas self-gravity, and general deviations from isothermality, and thus some caution must be taken when applying our results to real AGN disks.
\subsection{Accretion}
Although the gravitational radii of embedded black holes were under-resolved by orders of magnitude in our simulations, our study achieved sufficient scale separation to observe that accretion onto each black hole does not occur in a spherically symmetric manner, but rather through an accretion disk.
Additionally, accretion rates onto the embedded binaries in our simulations are about $\sim8\%$ of the Bondi accretion rate $\dot{M}_B=4\pi\rho_0G^2m_b^2c_s^{-3},$ which is reasonable given that our simulations are vertically stratified rather than uniform in density, and that the binary Hill radius is comparable to the Bondi radius $R_B=Gm_b/c_s^2$ such that tidal forces may limit accretion onto the binary. This is largely consistent with the results of previous three-dimensional simulations of accretion onto embedded binaries \citep{2022ApJ...940..155D}, but as noted in Section \ref{sec:large}, retrograde binaries accrete at a slightly higher rate than prograde ones due to their disparate gravitational influence on gas entering the Hill sphere. 

However, for typical gas densities in AGN disks such as those modeled in Figure \ref{fig:disks}, $\rho_0\sim10^{-18}-10^{-9}\,{\rm g\,cm^{-3}}$, the Bondi accretion rate may be substantially larger than the Eddington accretion rate. Fixing $R_H/H,$ as in our simulations, the Bondi accretion rate can be expressed as 
\begin{equation}
\begin{split}
\dot{M}_B=12\pi\left(\frac{R_H}{H}\right)^3\rho_0Gm_b\Omega^{-1}\approx1.18\left(\frac{R_H}{H}\right)^3
\left(\frac{m_b}{10\,M_\odot}\right)\\
\left(\frac{\rho_0}{10^{-14}\,\rm{g\,cm^{-3}}}\right)
\left(\frac{M_\bullet}{10^7\,M_\odot}\right)^{-1/2}\left(\frac{a_\bullet}{1\,\rm{pc}}\right)^{3/2}\,\rm{M_\odot\,yr^{-1}}.
\end{split}
\end{equation}

We can compare this to the Eddington-limited accretion rate onto a binary of the same mass,
\begin{equation}
\dot{M}_{\rm Edd} = \frac{4\pi Gm_b}{\kappa c \epsilon} \approx 2.2\times10^{-7}\left(\frac{0.1}{\epsilon}\right) \left(\frac{m_b}{10\,M_\odot}\right)\,\rm{M_\odot\,yr^{-1}},
\end{equation}
where $\epsilon$ is the fraction of rest mass radiated during accretion onto the black holes, $c$ is the speed of light, and
$\kappa$ is the opacity of the ambient gas. For brevity, we assume that $\kappa=0.4\,\rm{cm^2\,g^{-1}}$, the value for fully ionized hydrogen gas. Then, the ratio of the Bondi accretion rate to the Eddington accretion rate is 

\begin{equation}
\begin{split}
\frac{\dot{M}_B}{\dot{M}_{\rm Edd}} \approx 5\times10^7 
\left(\frac{R_H}{H}\right)^3\left(\frac{\rho_0}{10^{-14}\,\rm{g\,cm^{-3}}}\right)
\\\left(\frac{M_\bullet}{10^7\,M_\odot}\right)^{-1/2}
\left(\frac{a_\bullet}{1 {\rm pc}}\right)^{3/2}\left(\frac{\epsilon}{0.1}\right).
\end{split}
\end{equation}

Thus, for a wide range of parameters, the accretion rates onto the binaries in our simulations, at $\sim8\%$ of the Bondi rate, 
are highly super-Eddington. If the accretion rate onto each black hole is radially constant, the accretion process onto these embedded black holes may lead to 
significant feedback to larger scales in the form of jets and winds. However, such feedback would be driven on scales 
many orders of magnitude smaller than the smallest resolved in our simulations, and will be the subject of future work. 
Furthermore, thermodynamic effects neglected in our isothermal simulations may suppress the accretion rate (possibly as $\dot{M}\propto r^{1/2}$, see \citealt{2023ApJ...946...26G}), resulting in a much smaller accretion rate near the event horizon. 

If either embedded black hole has an appreciable spin, jets will likely be launched into the AGN disk. Because jet efficiency strongly depends on the angular momentum of the gas accreting onto a given black hole \citep[e.g.][]{2023ApJ...946L..42K}, it seems probable that prograde binaries could support jets; on the other hand, because the CSD size of retrograde binaries appears to depend strongly on softening, if they exist at all \citep{2022MNRAS.517.1602L}, retrograde binaries may not be able to launch strong jets.
Although on small scales jets are typically aligned with the spin of the black hole by which they are launched, 
on larger scales jets can be aligned with the accretion disk surrounding them \citep{2013Sci...339...49M,2018MNRAS.474L..81L}.
It remains to-be-determined whether jets launched by embedded black holes would reach scales comparable to the size of their minidisks, although if they do they may be re-collimated and oriented along the normal to the AGN disk midplane. 

\subsection{Orbital Evolution}\label{sec:discorb}
The two most robust results, in terms of orbital evolution, of our study of stellar-mass black hole binaries embedded in AGN disks are that retrograde binaries inspiral more quickly
and that binaries are gradually aligned AGN disk midplane. Specifically, anti-aligned binaries
inspiral more than four times faster than aligned binaries. Furthermore, the rate at which binary inclinations decreases depends
on how far out of the AGN disk midplane their orbital planes lie. Thus, binaries which form in a high-inclination retrograde configuration will both have their inclinations reduced and have their semi-major axes rapidly decreased through interactions with gas. 
Recent studies \citep[e.g.][]{2023ApJ...944L..42L} have suggested that a large fraction of binaries should form in retrograde configurations. 

Additionally, we have observed that both $\omega_b$ and $\Omega_b$ precess due to interactions with the AGN disk. Although the former is not well-defined  due to the low eccentricities in our simulations, and the latter is generally small, depending on the ambient gas density, 
the precession of these angles may affect various resonances which could operate on embedded binaries. It is well-known that 
precession of the argument of periapsis can both prevent ZLK cycles \citep{1997Natur.386..254H} and decrease the maximum achievable eccentricities 
in cases where ZLK cycles still occur \citep{2002ApJ...576..894M}. However, because of the small binary eccentricities in our simulations, the degree to 
which ZLK cycles will be affected is uncertain, although our results suggest that ZLK cycles should be less common for embedded binaries 
than those in vacuum. Evection resonances are more complicated still, due to their dependence on $n_\bullet$ as well as $\dot{\varpi}_b$. Interactions with the gas disk may either help or hinder evection resonances, although our low-$i_b$, low-$e_b$ simulations are 
not sufficient to predict how evection resonances should be affected in a general sense. 

The comparative rates of change of the semi-major axis and inclination aside, we expect that all binaries with inclinations $i_b<180^\circ$ will be gradually driven towards $i_b=0^\circ$. Thus, our results suggest that many retrograde binaries will pass through the range of inclinations where ZLK cycles occur as their inclinations are gradually damped. Although the rate of orbital evolution depends strongly on disk properties, if it is slow compared to the timescale of ZLK cycles, the binary will undergo ZLK oscillations in inclination and eccentricity as long as they are not disrupted by disk-induced precession. Moreover, binaries near $i_b\sim90^\circ$ may approach $e_b\sim1$ via ZLK oscillations, in which case the dissipation of energy and angular momentum due to gravitational waves could become extreme near pericentre leading to a large population of merging black holes with polar alignment relative to the AGN disk.

\section{Conclusions}\label{sec:conclusion}

We have conducted a series of three-dimensional hydrodynamical simulations of stellar-mass binaries embedded in AGN disks. Whereas previous three-dimensional studies of embedded binaries focused on those with prograde orbits aligned with the midplane of the AGN disk, we have explored a range of binary inclinations including both prograde and retrograde configurations. Our work has uncovered qualitatively new aspects to both the accretion flows around and orbital evolution of inclined embedded binaries. However, we caution that our simulations have used a simplified isothermal equation of state, neglected magnetic fields, and focused on nearly-circular binaries misaligned by not more than $30^\circ$ from the AGN disk midplane.

Binaries with higher inclinations tend to accrete at higher rates, in all cases accreting at less than the Bondi rate. Retrograde binaries have substantially smaller minidisks, although the black holes constituting the $i_b=180^\circ$ binary possessed larger minidisks than those constituting the $i_b=150^\circ$ or $i_b=165^\circ$ binaries, due at least in part to ram pressure stripping. 
We find that the minidisks around each black hole are typically aligned with the AGN disk midplane.
On the larger scales of the minidisks and AGN disk there are patterns of midplane outflows, meridional circulation, and accretion along the surface of gaseous tori. Some gas also flows directly onto each black hole along the polar directions. 

Retrograde binaries inspiral up to four times as quickly as their prograde counterparts due to gravitational interactions with the disk.
Additionally, we find that binaries with orbital planes outside of the AGN disk midplane experience inclination damping, which is small for binaries only slightly misaligned with the disk midplane, but grows larger for more misaligned binaries. We have also found that eccentricities can be excited in near-circular inclined binaries, although we cannot comment on the maximum eccentricities achievable through disk-binary interactions due to the limited range of orbital parameters surveyed herein. Similarly, we have found evidence for disk-driven precession of both the binary argument of periapsis and longitude of the ascending node, which may affect various the susceptibility of embedded binaries to various resonances. Because a significant fraction of binaries in AGN disks may form with retrograde configurations, our results paint an optimistic, albeit incomplete, picture of gas interactions assisting binary inspirals down to to scales where gravitational waves can drive their merger, resulting in gravitational wave signals such as those observed by LIGO/VIRGO. 

\section*{Software}

\texttt{matplotlib} \citep{4160265}, \texttt{cmocean} \citep{cmocean}, \texttt{numpy} \citep{5725236}, \texttt{yt} \citep{2011ApJS..192....9T}, \texttt{Athena++} \citep{2020ApJS..249....4S}, \texttt{REBOUND} \citep{2012A&A...537A.128R,2015MNRAS.446.1424R}

\section*{Acknowledgments}
We are particularly grateful for discussions with and feedback from Josh Calcino, and AJD is thankful for numerous discussions on gravitational dynamics with Doug Hamilton.
We gratefully acknowledge the support by LANL/LDRD under project number 20220087DR. The authors acknowledge the University of Maryland supercomputing resources (http://hpcc.umd.edu) made available for conducting the research reported in this paper. This research used resources provided by the Los Alamos National Laboratory Institutional Computing Program, which is supported by the U.S. Department of Energy National Nuclear Security Administration under Contract No. 89233218CNA000001. The LA-UR number is LA-UR-22-33104.

\bibliographystyle{aasjournalnolink}
\bibliography{references}

\appendix

\section{Orbital evolution diagnostics}\label{sec:appendix}
We detail hereinafter how we measure changes in binary properties such as angular momentum and energy; and how we use these quantities to calculate the average rates of changes of binary orbital elements reported in Section \ref{sec:orbits}.
\subsection{Evolution equations}
Expressions for the time evolution of the binary semi-major axis and eccentricity in terms of changes in the mass, energy, and angular momentum of the binary are given by Equations (\ref{eq:abdot}) and (\ref{eq:ebdot}) respectively. Equations (\ref{eq:idot}), (\ref{eq:Omdot}), and (\ref{eq:omegadot}) provide expressions for the rate of change of the binary inclination, longitude the ascending node, and argument of periapsis respectively. For completeness, we note that
\begin{align}
\frac{\dot{\mu}_b}{\mu_b} = \frac{\dot{m}_1}{m_1}+\frac{\dot{m}_2}{m_2}-\frac{\dot{m}_b}{m_b},\\
\frac{\dot{J}_i}{J_i} = \frac{\dot{\mu}_b}{\mu_b} + \frac{\dot{h}_i}{h_i},\\
\frac{\dot{E}}{E} = \frac{\dot{\mu}_b}{\mu_b} + \frac{\dot{\mathcal{E}}}{\mathcal{E}},
\end{align}
and 
\begin{align}
\dot{\mathcal{E}} = \mathbf{a}_{\rm ext}\cdot\mathbf{v} - \frac{G\dot{m}_b}{|\mathbf{r}|},\\
\dot{\mathbf{h}} = \mathbf{r}\times\mathbf{a}_{\rm ext},\\
\dot{\mathbf{e}}= \frac{1}{Gm_b}\left(\mathbf{a}_{\rm ext}\times\mathbf{h} + \mathbf{v}\times\dot{\mathbf{h}} - (\mathbf{v}\times\mathbf{h})\frac{\dot{m}_b}{m_b}\right),
\end{align}
where we have assumed a Keplerian binary orbit perturbed by external accelerations $\mathbf{a}_{\rm ext}$ and specific torques $\dot{\mathbf{h}}$.

\subsection{Simulation diagnostics}\label{sec:diagnostics}
Every time the point masses in our simulations interact with the fluid through source terms (when using Runge-Kutta time integration schemes, as in this work, this simply occurs every substep), the point masses accrete some gas from within the sink region and gravitationally interact with the surrounding fluid. The gravitational force on each object ($\mathbf{F}_{g,i}$) is given by
\begin{equation}
\mathbf{F}_{g,i} = \int \rho \nabla\Phi_i dV,
\end{equation}
and the analogous contribution due to accretion, following from the sink term $\mathbf{S}_p$ in Equations (\ref{eq:mass}) and (\ref{eq:momentum}), is
\begin{equation}
\mathbf{F}_{a,i} = -\int \mathbf{S}_{p,i} dV,
\end{equation}
Similarly, the accretion rate onto the binary $\dot{m}_b$ is simply
\begin{equation}
\dot{m}_b=-\int S_\Sigma dV.
\end{equation}
After measuring the force on each point mass, we can calculate their accelerations as
$\mathbf{a}_{a, \rm ext}=(\mathbf{F}_{a,2} - \mathbf{v}_2\dot{m}_2)/m_2 - ((\mathbf{F}_{a,1} - \mathbf{v}_1\dot{m}_1)/m_1)$ and 
$\mathbf{a}_{g, \rm ext}=\mathbf{F}_{g,2}/m_2 - \mathbf{F}_{g,1}/m_1$. 

\subsection{Averaging over time}\label{sec:averaging}
Although we averaged the force on each particle over every time step, we record derived quantities discretely using $452$ samples per binary orbit. 
Measuring changes in orbital elements is complicated by the fact that even in the absence of feedback from the gas during accretion or through gravity, the orbital elements of the binary vary over time due to 3-body interactions with the SMBH (see, e.g. Figure \ref{fig:orbit165}). It is straightforward to calculate the average of a quantity $Q$ in time as 
\begin{equation}
\langle Q \rangle=\frac{1}{t_2-t_1}\int_{t_1}^{t_2}Qdt.
\end{equation}

For example, one of our aims is to measure a characteristic rate of change of the binary semi-major axis, $\langle \dot{a}_b/a_b\rangle$. Instantaneously, it is straightforward to measure the change in semi-major axis in terms of the binary energy and masses of the binary components as 
\begin{equation}
\frac{d\log{a_b}}{d\log{m_b}}=\frac{\dot{a}_b}{a_b}\frac{m_b}{\dot{m}_b}=\left(\frac{\dot{m}_1}{m_1}+\frac{\dot{m}_2}{m_2}-\frac{\dot{E}}{E}\right)\frac{m_b}{\dot{m}_b}
\end{equation}
However, there is some ambiguity because $a_b$ and $E_b$ evolve in time through tidal interactions with the SMBH. Thus, instead of reporting $\langle \dot{a}_b/a_b\rangle$, we report quantities like $\langle \dot{a}_b \rangle/\langle a_b \rangle$, additionally normalized by the accretion timescale $m_b/\langle \dot{m}_b \rangle$.

\section{The Influence of Gravitational Softening}\label{sec:appendixSoft}
As mentioned in Section \ref{sec:minidisks}, we found that the size of the minidisks around each of the black holes in retrograde binaries were not converged, in the sense that the size of those CSDs apprear to depend on length scale for our sink and gravitational softening regions. In the main body of the text, we reported results using a spline softening length of $0.04a_b$.\footnote{Because spline softening reduces to an exact Keplerian potential outside of the sink region, a roughly commensurate Plummer softening would have a softening length $\sim2.8$ times larger \citep[e.g.][]{2001NewA....6...79S}.} 

\begin{figure}
\includegraphics[width=\linewidth]{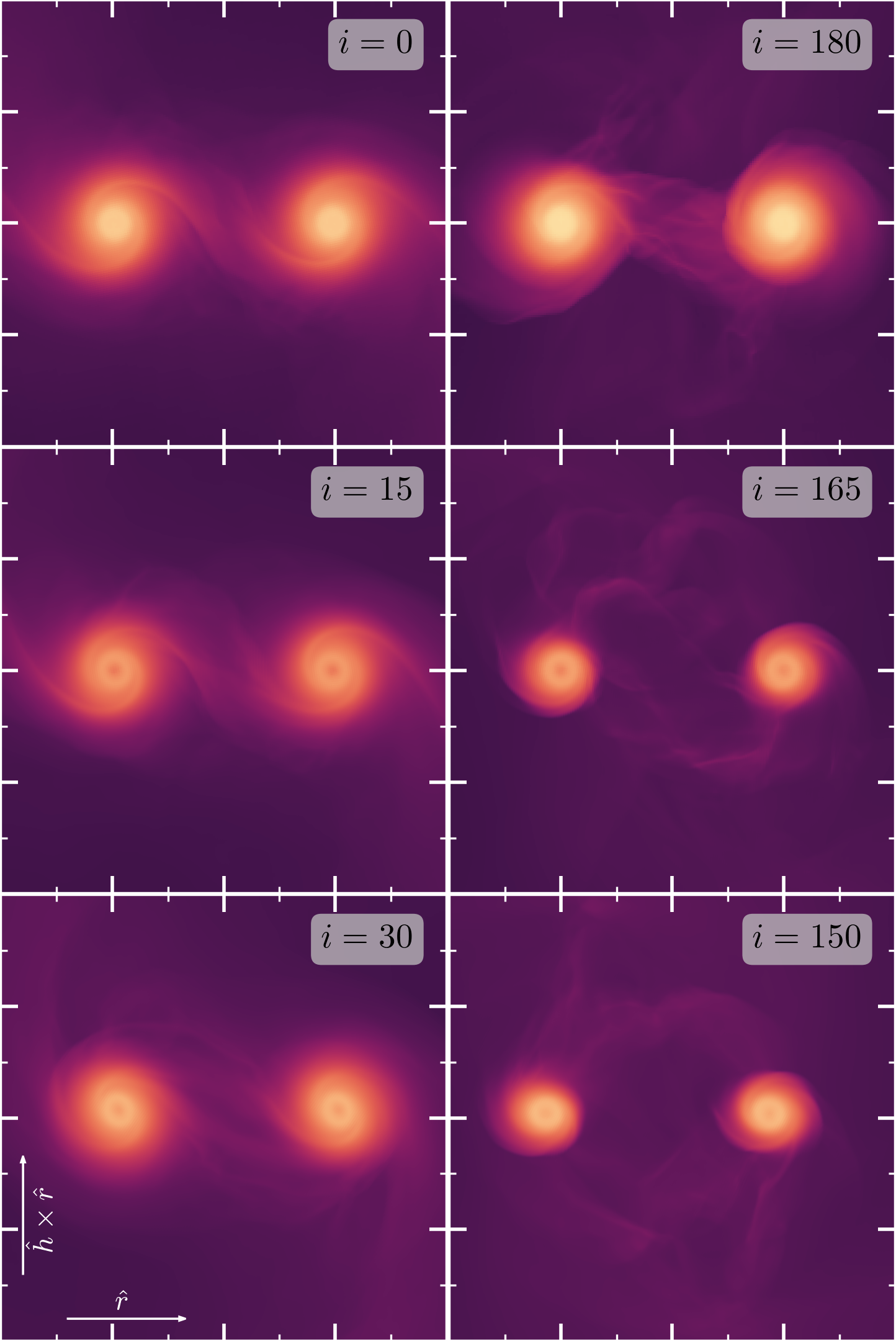}
\caption{Projections of the fluid density along the orbital angular momentum vector of each binary at $t=18.4n_\bullet^{-1}.$, in analogy to Figure \ref{fig:projecting}. However, each point mass in the simulations pictured here has a gravitational softening length (and sink length) of $0.12a_b$, three times larger than the simulations presented in the main text. Many features on larger scales are completely consistent between these simulations, such as the streams of gas being stripped off of the retrograde binaries and gas flowing between the members of the prograde binaries. However, in these simulations the minidisks around the black holes in retrograde binaries are appreciably larger than in simulations using smaller softening lengths. Major ticks are placed $0.5a_b$ apart, with minor ticks every $0.25a_b$.}\label{fig:project12}
\end{figure}

To illustrate how CSD size depends on the softening length, we plot in Figure \ref{fig:project12} projections of the gas density along the binary angular momentum vector analogous to Figure \ref{fig:projecting}, except for simulations using sink and softening lengths of $0.12a_b$, the same settings as \citet{2022ApJ...940..155D}. The minidisks are virtually unchanged for prograde binaries. However, the CSDs of retrograde binaries are visibly larger in Figure \ref{fig:project12} than in Figure \ref{fig:projecting}; the minidisks in the $i_b=180$ case are about the same size as those in the $i_b=0$ case, and the minidisks of the $i_b=165$ and $i_b=150$ binaries are both about the same size as each other and appreciably larger than their smaller-softening counterparts.

\begin{figure}
\includegraphics[width=\linewidth]{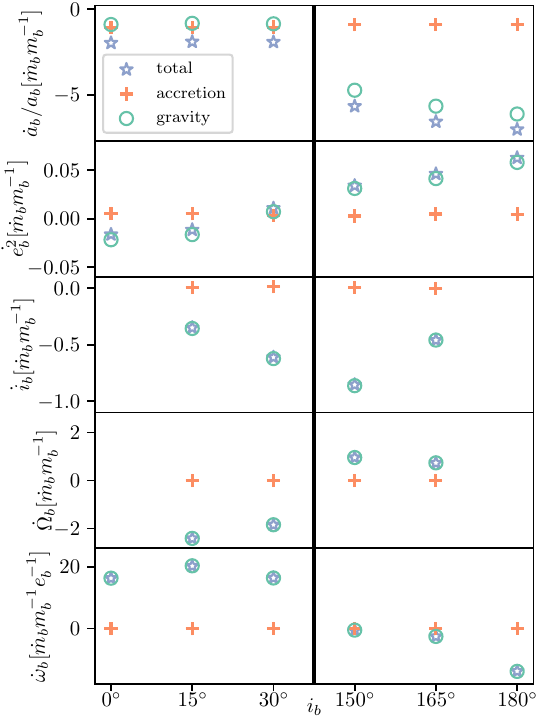}
\caption{The same quantities shown in Figure \ref{fig:orbital} for our simulations with the same resolution but gravitational softening parameters and sink rates twice as large, in this case $0.08a_b$ compared to the value of $0.04a_b$ used in the body of the paper.}
\label{fig:orbital8}
\end{figure}

\begin{deluxetable*}{ccccccccc}
\caption{Time-averaged values relevant to binary orbital evolution.}\label{tab:orbit}
\tablehead{
\colhead{$i_b$} & \colhead{$\frac{\langle\dot{m}_b\rangle}{\dot{M}_B}$} & \colhead{$\left.\frac{\langle\dot{a}_b\rangle m_b}{\langle a_b\rangle \langle\dot{m}_b\rangle}\right.$} & \colhead{$\left.\frac{\langle\dot{a}_b\rangle m_b}{\langle a_b\rangle \langle\dot{m}_b\rangle}\right|_{\rm grav.}$} & \colhead{$\left.\frac{\langle\dot{e}^2_b\rangle m_b}{\langle\dot{m}_b\rangle}\right.$} & \colhead{$\left.\frac{\langle\dot{i}_b\rangle m_b}{\langle\dot{m}_b\rangle}\right.$} & \colhead{$\left.\frac{\langle\dot{\Omega}_b\rangle m_b}{\langle\dot{m}_b\rangle}\right.$} & \colhead{$\left.\frac{\langle\dot{\omega}_b\rangle m_b}{\langle e_b\rangle\langle\dot{m}_b\rangle}\right.$} & \colhead{$\langle e_b\rangle$}
}
\startdata
$0^\circ$   &0.0712 & -1.76 & -0.634 & -0.0183 & - & - & 10.8 & 0.0157 \\
$15^\circ$  &0.0722 & -1.74 & -0.671 & -0.0126 & -0.292 & -0.237 & 14.2 & 0.0155 \\
$30^\circ$  &0.0725 & -1.89 & -0.842 & -0.0092 & -0.508 & -0.329 & 10.4 & 0.0155\\
$150^\circ$ &0.0844 & -5.71 & -4.80  & 0.0318  & -0.868 & 0.249 & -0.419 & 0.0170\\
$165^\circ$ &0.0873 & -6.55 & -5.65  & 0.056   & -0.459 & 0.0316 & 0.09349 & 0.0134 \\
$180^\circ$ &0.0863 & -7.12 & -6.21  & 0.0572  & - & - & -5.15 & 0.0131 \\
\enddata
\end{deluxetable*}

Although the sizes of the minidisks appear to differ based on the gravitational softening length, we have found that this results in only very minor changes in binary orbital evolution. To illustrate this, we have plotted the time-averaged rates of change of the binary orbital elements from a suite of simulations using a softening length of $0.08a_b$, twice that used in the simulations comprising the main body of our study. The rates of change of the binary semi-major axis, eccentricity, and inclination are virtually identical to the results from smaller-softening simulations. However, we find that the time-averaged values of $\dot{\Omega}_b$ and $\dot{\omega}_b$ differ by factors of order unity between the $0.04a_b$- and $0.08a_b$-softening simulations, though in all cases these fluid-induced precession rates could hinder ZLK cycles. 

Notably, the $i_b=0$ simulations suggest an appreciably slower rate of inspiral for binaries with $a_b/R_h=1/4$ binaries than the equivalent simulations in \citet{2022ApJ...940..155D}: $\langle \dot{a}_b \rangle/\langle a_b\rangle \approx -1.76 \langle\dot{m}_b\rangle/m_b$ in our simulations compared to $\langle \dot{a}_b \rangle/\langle a_b\rangle \approx -3.1 \langle\dot{m}_b\rangle/m_b$ in \citet{2022ApJ...940..155D}. However, \citet{2022ApJ...940..155D} used not only a softening length of $0.12a_b$ in comparison to the default value of $0.04a_b$ in this work, but also a different method of solving the equations of motion of the binary. Thus, we have repeated our $i_b=0$ simulation using softening lengths of $0.12a_b,~0.08a_b,~0.04a_b,$ and $0.02a_b$, using the same value for the sink length in each case. All of these simulations used our standard resolution settings except for the smallest-softening simulations, for which we added an additional level of refinement around the binary to achieve a resolution of $\sim2400$ cells per $a_b$. Because of the great cost associated with the highest-resolution simulation, it was only carried out until $t=5.5n_\bullet^{-1}$. Although not enough to reach a converged state or achieve robust statistics, this was sufficient for the binary to reach a quasi-steady accretion rate. Because the accretion contribution to the evolution of the semi-major axis is independent of softening length, we focus on the gravitational contribution to $\langle\dot{\mathcal{E}}\rangle/\langle\mathcal{E}\rangle$. For the aforementioned softening lengths, we found values of $\langle\dot{\mathcal{E}}\rangle/\langle\mathcal{E}\rangle$, averaged from $t=4.5n_\bullet^{-1}$ to $t=5.5n_\bullet^{-1}$, of $\sim2.5,0.78,0.57,0.76$. Thus, it appears that our fiducial softening choice is sufficiently converged, at least for prograde binaries, although our results can be assumed uncertain at the level of a few percent.

\section{Quantitative Records of orbital evolution}\label{sec:appendixTable}
We list in Table \ref{tab:orbit} a number of the quantities plotted in Figure \ref{fig:orbital}, as well as the accretion rate onto each binary, in all cases averaged from $t=14n_\bullet^{-1}$ to $t=20n^{-1}_\bullet$.

\end{document}